\documentclass[fleqn,usenatbib]{mnras}

\usepackage{newtxtext,newtxmath, comment}

\usepackage[T1]{fontenc}

\DeclareRobustCommand{\VAN}[3]{#2}
\let\VANthebibliography\thebibliography
\def\thebibliography{\DeclareRobustCommand{\VAN}[3]{##3}\VANthebibliography}

\usepackage{graphicx}	
\usepackage{amsmath}	
\usepackage{cuted}
\usepackage{multirow} 
\usepackage{nccmath}
\numberwithin{equation}{section}

\title[Mutual Coupling in HERA]{Investigating Mutual Coupling in the Hydrogen Epoch of Reionization Array and Mitigating its Effects on the 21-cm Power Spectrum}

\author[E.\ Rath\ \&\ R.\ Pascua et al.]{
        E. Rath$^{1,2}$\thanks{Both authors contributed equally to this work.}\thanks{Email: elerath@mit.edu},
        R. Pascua$^{3}$\footnotemark[1]\thanks{Email: robert.pascua@mail.mcgill.ca},
        A. T. Josaitis$^{4}$,
        A. Ewall-Wice$^{5}$,
        N. Fagnoni$^{4}$,
        E. de Lera Acedo$^{4,6}$,
    \newauthor
        Z. E. Martinot$^{7}$,
        Z. Abdurashidova$^{8}$,
        T. Adams$^{9}$,
        J. E. Aguirre$^{7}$,
        R. Baartman$^{9}$,
        A. P. Beardsley$^{10}$,
    \newauthor
        L. M. Berkhout$^{11}$,
        G. Bernardi$^{9,12,13}$,
        T. S. Billings$^{7}$,
        J. D. Bowman$^{11}$,
        P. Bull$^{14,19}$,
        J. Burba$^{14}$,
    \newauthor
        R. Byrne$^{15}$,
        S. Carey$^{4}$,
        K.-F. Chen$^{1,2}$,
        S. Choudhuri$^{16}$,
        T. Cox$^{8}$,
        D. R. DeBoer$^{17}$,
        M. Dexter$^{17}$,
    \newauthor
        J. S. Dillon$^{8}$,
        S. Dynes$^{1}$,
        N. Eksteen$^{9}$,
        J. Ely$^{4}$,
        R. Fritz$^{9}$,
        S. R. Furlanetto$^{18}$,
        K. Gale-Sides$^{4}$,
    \newauthor
        H. Garsden$^{14}$,
        B. K. Gehlot$^{11}$,
        A. Ghosh$^{19}$,
        A. Gorce$^{3,20}$,
        D. Gorthi$^{8}$,
        Z. Halday$^{9}$,
        B. J. Hazelton$^{15,21}$,
    \newauthor
        J. N. Hewitt$^{1,2}$,
        J. Hickish$^{17}$,
        T. Huang$^{4}$,
        D. C. Jacobs$^{11}$,
        N. S. Kern$^{2,\parallel}$,
        J. Kerrigan$^{22}$,
        P. Kittiwisit$^{19}$,
    \newauthor
        M. Kolopanis$^{11}$,
        A. Lanman$^{22}$,
        A. Liu$^{3}$,
        Y.-Z. Ma$^{23}$,
        D. H. E. MacMahon$^{17}$,
        L. Malan$^{9}$,
        C. Malgas$^{9}$,
    \newauthor
        K. Malgas$^{9}$,
        B. Marero$^{9}$,
        L. McBride$^{3,20}$,
        A. Mesinger$^{24}$,
        N. Mohamed-Hinds$^{15}$,
        M. Molewa$^{9}$,
    \newauthor
        M. F. Morales$^{15}$,
        S. G. Murray$^{11,24}$,
        B. Nikolic$^{4}$,
        H. Nuwegeld$^{9}$,
        A. R. Parsons$^{8}$,
        N. Patra$^{8}$,
    \newauthor
        P. La Plante$^{25,26}$,
        Y. Qin$^{27}$,
        N. Razavi-Ghods$^{4}$,
        D. Riley$^{1}$,
        J. Robnett$^{28}$,
        K. Rosie$^{29}$,
        M. G. Santos$^{9,19}$,
    \newauthor
        P. Sims$^{11}$,
        S. Singh$^{3}$,
        D. Storer$^{15}$,
        H. Swarts$^{9}$,
        J. Tan$^{7}$,
        M. J. Wilensky$^{14}$,
        P. K. G. Williams$^{30,31}$,
    \newauthor
        P. van Wyngaarden$^{9}$,
        H. Zheng$^{2}$
\\
    $^{1}$ MIT Kavli Institute, Massachusetts Institute of Technology, Cambridge, MA\\
    $^{2}$ Department of Physics, Massachusetts Institute of Technology, Cambridge, MA\\
    $^{3}$ Department of Physics and Trottier Space Institute, McGill University, 3600 University Street, Montreal, QC H3A 2T8, Canada\\
    $^{4}$ Cavendish Astrophysics, University of Cambridge, Cambridge, UK\\
    $^{5}$ Department of Physics, University of California, Berkeley, CA\\
    $^{6}$ Kavli Institute for Cosmology in Cambridge, University of Cambridge, Cambridge, UK\\
    $^{7}$ Department of Physics and Astronomy, University of Pennsylvania, Philadelphia, PA\\
    $^{8}$ Department of Astronomy, University of California, Berkeley, CA\\
    $^{9}$ South African Radio Astronomy Observatory, Black River Park, 2 Fir Street, Observatory, Cape Town, 7925, South Africa\\
    $^{10}$ Department of Physics, Winona State University, Winona, MN\\
    $^{11}$ School of Earth and Space Exploration, Arizona State University, Tempe, AZ\\
    $^{12}$ INAF-Istituto di Radioastronomia, via Gobetti 101, 40129 Bologna, Italy\\
    $^{13}$ Department of Physics and Electronics, Rhodes University, PO Box 94, Grahamstown, 6140, South Africa\\
    $^{14}$ Jodrell Bank Centre for Astrophysics, University of Manchester, Manchester, M13 9PL, United Kingdom\\
    $^{15}$ Department of Physics, University of Washington, Seattle, WA\\
    $^{16}$ Centre for Strings, Gravitation and Cosmology, Department of Physics, Indian Institute of Technology Madras, Chennai 600036, India\\
    $^{17}$ Radio Astronomy Lab, University of California, Berkeley, CA\\
    $^{18}$ Department of Physics and Astronomy, University of California, Los Angeles, CA\\
    $^{19}$ Department of Physics and Astronomy,  University of Western Cape, Cape Town, 7535, South Africa\\
    $^{20}$ Institut d’Astrophysique Spatiale, CNRS, Université Paris-Saclay, 91405 Orsay, France\\
    $^{21}$ eScience Institute, University of Washington, Seattle, WA\\
    $^{22}$ Department of Physics, Brown University, Providence, RI\\
    $^{23}$ Department of Physics, Stellenbosch University, Stellenbosch, South Africa\\
    $^{24}$ Scuola Normale Superiore, 56126 Pisa, PI, Italy\\
    $^{25}$ Department of Computer Science, University of Nevada, Las Vegas, NV\\
    $^{26}$ Nevada Center for Astrophysics, University of Nevada, Las Vegas, NV\\
    $^{27}$ School of Physics, University of Melbourne, Parkville, VIC 3010 Australia\\
    $^{28}$ National Radio Astronomy Observatory, Socorro, NM 87801, USA\\
    $^{29}$ South African Astronomical Observatory, Cape Town, 7925, South Africa\\
    $^{30}$ Center for Astrophysics, Harvard \& Smithsonian, Cambridge, MA\\
    $^{31}$ American Astronomical Society, Washington, DC\\
    $^{\parallel}$ NASA Hubble Fellow\\
}

\date{Accepted XXX. Received YYY; in original form ZZZ}
\pubyear{2024}
\begin{document}
\label{firstpage}
\pagerange{\pageref{firstpage}--\pageref{lastpage}}
\maketitle

\begin{abstract}
Interferometric experiments designed to detect the highly redshifted 21-cm signal from neutral hydrogen are producing increasingly stringent constraints on the 21-cm power spectrum, but some $k$-modes remain systematics-dominated.
Mutual coupling is a major systematic that must be overcome in order to detect the 21-cm signal, and simulations that reproduce effects seen in the data can guide strategies for mitigating mutual coupling.
In this paper, we analyse 12 nights of data from the Hydrogen Epoch of Reionization Array and compare the data against simulations that include a computationally efficient and physically motivated semi-analytic treatment of mutual coupling.
We find that simulated coupling features qualitatively agree with coupling features in the data; however, coupling features in the data are brighter than the simulated features, indicating the presence of additional coupling mechanisms not captured by our model.
We explore the use of fringe-rate filters as mutual coupling mitigation tools and use our simulations to investigate the effects of mutual coupling on a simulated cosmological 21-cm power spectrum in a `worst case' scenario where the foregrounds are particularly bright.
We find that mutual coupling contaminates a large portion of the `EoR Window', and the contamination is several orders-of-magnitude larger than our simulated cosmic signal across a wide range of cosmological Fourier modes.
While our fiducial fringe-rate filtering strategy reduces mutual coupling by roughly a factor of 100 in power, a non-negligible amount of coupling cannot be excised with fringe-rate filters, so more sophisticated mitigation strategies are required. 
\end{abstract}



\section{Introduction} \label{sec:INTRO}
A new generation of low frequency interferometers promise significant advances in many fields of astronomy \citep{Newburgh:2016, HERA_overview, Hallinan:2019, McKinnon:2019, Vanderlinde:2019}.
One of the primary science goals of these current and next-generation radio interferometers is to detect the cosmologically redshifted 21-cm signal from Cosmic Dawn (CD) and the Epoch of Reionization (EoR), the period of cosmic history when the first stars and galaxies illuminated and reionized the universe.
Detections of the cosmological 21-cm signal will provide valuable insights into the nature of the first stars and galaxies and will shed light on the details of how reionization proceeded (see, e.g., \citealt{21cm_Van_de_Hulst_1945,Field1959,21cm_general_Rees_Madau}; and \citealt{Wang_2006}).
A number of pathfinder experiments, including the Murchison Widefield Array (MWA, \citealt{MWA_overview}), the Low Frequency Array (LOFAR, \citealt{LOFAR_overview}), and the Hydrogen Epoch of Reionization Array (HERA, \citealt{HERA_overview}), are aiming for a first detection of the cosmological 21-cm power spectrum during CD and the EoR.
These experiments are currently operational and producing increasingly sensitive upper-limits on the 21-cm power spectrum \citep{MWA_2023_Limits,HERA_2023_Limits,LOFAR_2020_Limits}.
Detecting the cosmological 21-cm signal, however, is full of challenges; reviews such as \citet{Furlanetto2006}, \citet{Pritchard_Loeb_Review}, \citet{Mesinger_Review}, and \citet{LiuShaw2020} discuss the experimental and theoretical challenges in measuring and interpreting the 21-cm signal. 

One of the greatest challenges in 21-cm cosmology lies in disentangling the 21-cm signal from the overwhelmingly bright astrophysical foregrounds, which are expected to be several orders of magnitude brighter than the cosmological signal.
One commonly employed approach, used by both the MWA and HERA, is the method of \emph{foreground avoidance}~\citep{Foreground_Wedge}, where the 21-cm signal is distinguished from the foregrounds by appealing to their differing spectral signatures.
Galactic and extra-galactic synchrotron radiation, which evolves smoothly in frequency, is the dominant emission mechanism for the foregrounds~\citep{Wang_2006}; this is contrasted against the fine-scale spectral fluctuations expected in the 21-cm signal due to the patchy nature of reionisation (e.g.,~\citealt{McQuinn_2006,Gnedin_2004}).
In principle, the delay spectrum estimator~\citep{Parsons_2012}, reviewed in Section~\ref{sec:PSPEC-ESTIMATION}, can separate the spectrally smooth foregrounds from the 21-cm signal, since the spectrally smooth component of the signal is confined to small cosmological Fourier modes.
In practice, this separation is difficult to achieve due to a variety of systematic effects, including anthropogenic Radio Frequency Interference (RFI), Faraday rotation by the ionosphere \citep{Jelic2010LOFAR_foregorunds, polarized_foregrounds, Polarized_ionosphere_attenuation}, circuit level cross-talk \citep{Zheng2014_MITEOR}, intra-antenna reflections and resonances \citep{Craeye_Macro_Basis_Func_Currents, Nithya2016_Blackman_Harris, Hibbard2020}, and calibration errors~\citep{Orosz_2019,Byrne_2019,Barry_2016}, among other effects.
These observational artefacts could introduce structure in the data similar to the 21-cm signal, so successful detection of the cosmological 21-cm signal depends on careful mitigation of these effects which could otherwise easily destroy or mask cosmological information. 

In addition to the challenges outlined above, systematic effects caused by over-the-air mutual coupling between antennas are a known challenge for measurements of the 21-cm power spectrum \citep{Systematics_Chaudhari, Kern:2019, Borg:2020, McKinley:2020, Bolli:2022}. 
All of the current upper limits on the 21-cm power spectrum report excesses beyond the expected noise level across a broad range of redshifts and primarily at low $k$, with mutual coupling often cited as one possible source of the excess power~\citep{PAPER_limit_2015, MWA_limit_2016_1, MWA_limit_2016_2, Kern:2019, Kern2020a, LOFAR_2020_Limits}.
Conclusively linking the observed excess power to mutual coupling requires a predictive model for how coupling effects manifest in the data, as well as comparisons of those predictions against what is observed in the data.

Fundamentally, a detailed understanding of mutual coupling requires knowledge of how radiation is reflected around the array and how that manifests as excess correlation between antennas.
Since these reflections occur at scales of a few wavelengths, numerical electromagnetic simulations are required to obtain a detailed understanding of mutual coupling; however, numerical artefacts between different electromagnetic solvers at high dynamic range~\citep{Mahesh_2021} and the intractable computational demands of running such solvers on large arrays of antennas makes this approach infeasible.
It is therefore necessary to use approximate, yet physically motivated, models to further our understanding of systematic features imparted on the data by mutual coupling.
\citet{Josaitis:2021a} proposed such a model of mutual coupling: an impedance mismatch at the input to an antenna's amplifier reflects the incident astrophysical signal, which is then re-radiated to and absorbed by other antennas in the interferometric array.
While this model does not capture all of the potential coupling mechanisms, it does provides a specific mechanism for mutual coupling that is computationally efficient and physically intuitive.
Using this model,~\citet{Josaitis:2021a} simulated mutual coupling for a variety of array configurations with HERA antennas and presented a phenomenological analysis of the coupling features that were produced.
In this paper, we expand on the work of~\citet{Josaitis:2021a} with a revised coupling model and improved simulation algorithm.
We confront these simulations with reality and assess the accuracy of the revised coupling model by comparing our simulations with data taken from the Phase II HERA instrument~\citep{Berkhout_2024}.
We additionally use our simulations to explore the efficacy of filtering as a method of mitigating mutual coupling and better understand the challenges mutual coupling poses for detecting the cosmological 21-cm power spectrum.


The structure of this paper is as follows:
In Section~\ref{sec:FORMALISM}, we provide an updated derivation of the first-order mutual coupling model and describe how the model is implemented in simulation;
In Section~\ref{sec:OBSERVATIONS}, we describe the HERA observations and visibility data used in this analysis;
In Section~\ref{sec:EVIDENCE}, we compare our simulations of mutual coupling against systematics seen in the data;
In Section~\ref{sec:MITIGATION}, we describe two filtering strategies designed to mitigate mutual coupling as well as the simulations used to test the efficacy of such filters;
In Section~\ref{sec:POWER-SPECTRUM}, we discuss the effects of mutual coupling on the estimated 21-cm power spectrum and provide a cursory investigation of how the proposed filtering strategies aid in detection of the cosmological 21-cm signal; 
In Section~\ref{sec:CONCLUSION} we summarise our results. 

\begin{figure*}
    \centering
    \includegraphics[width=\textwidth]{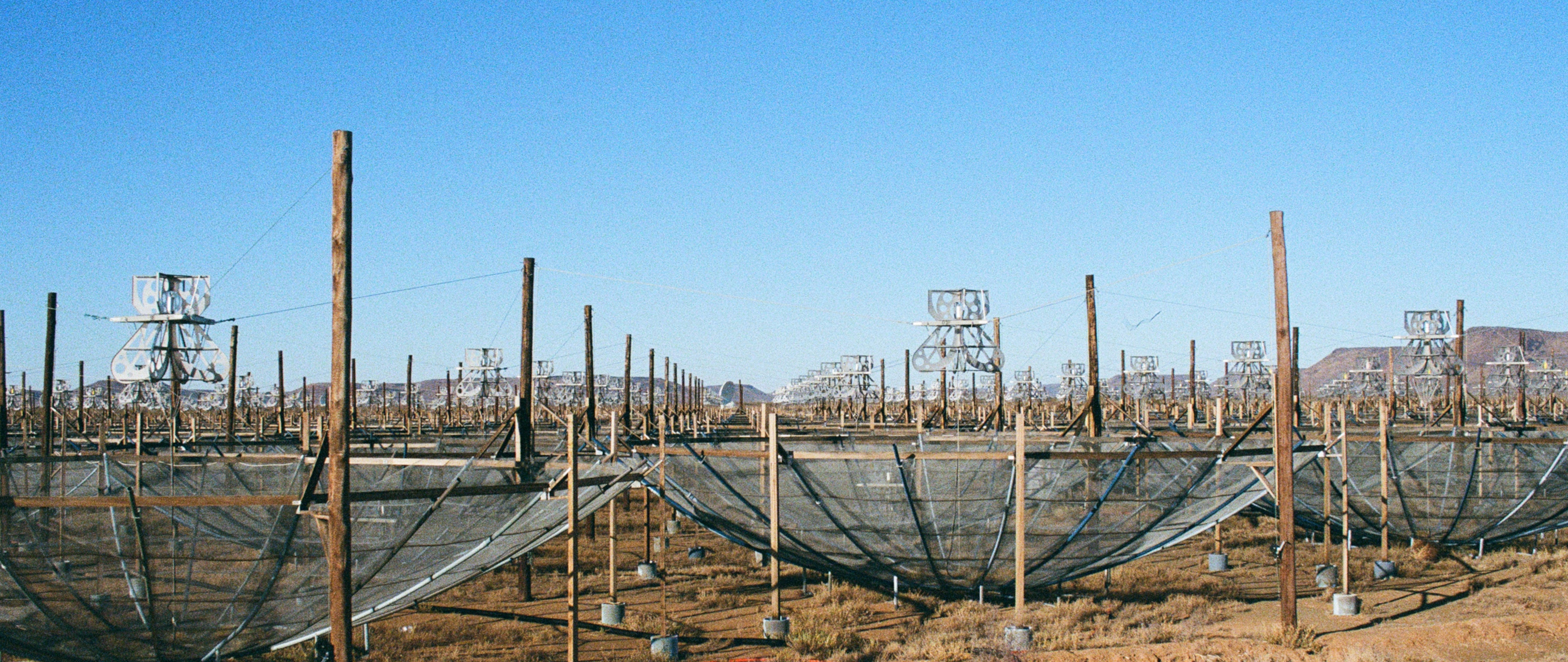}
    \caption{
        Photo of the Phase II HERA antennas.
        This view highlights the fact that many pairs of feeds have unobstructed views of each other and motivates the feed-to-feed re-radiation model of mutual coupling presented in Section~\ref{sec:FORMALISM}.
    }
    \label{fig:array_photo}
\end{figure*}

\section{Revisiting First-Order Coupling}
\label{sec:FORMALISM}

Before moving forward, we want to clearly define the context of this work as well as a few terms which have broad colloquial meanings.
Our analysis is restricted to zenith-pointing, drift-scan interferometric arrays which do not track particular positions on the sky.
Instead, these arrays focus their sensitivity directly overhead and achieve different `pointings' through Earth rotation.
When referring to an \emph{antenna}, we are referring to the combination of the dish (the parabolic reflecting surface in~\autoref{fig:array_photo}) with the feed that is suspended above the centre of the dish.
We may use the terms `antenna' and `array element' interchangeably.

In this section, we revisit the semi-analytic model of mutual coupling proposed by~\citet{Josaitis:2021a} to provide an updated expression for the `first-order' coupled visibilities.
Their model is `semi-analytic' in the sense that numerical electromagnetic simulations are used to determine the impedance and the radiation field for an isolated antenna, and then the outputs from the numerical simulations are used to evaluate the results of an analytic treatment of the re-radiation coupling mechanism.
Performing the analytic calculation reveals that the coupled visibilities may be expressed in terms of uncoupled quantities involving the isolated antenna properties and the `zeroth-order' visibilities that would be measured in the absence of any coupling mechanisms.
The isolated antenna properties may be absorbed into effective reflection (or coupling) coefficients, and the `first-order' model is obtained by keeping only terms involving at most one reflection coefficient (or, in other words, by only considering `single-scattering' events).

~\autoref{fig:array_photo} provides a relatively clear physical motivation for the model proposed by~\citet{Josaitis:2021a}: each antenna's feed has an unobstructed view of many other antennas' feeds.
Therefore, if an antenna were to re-radiate some of the incident astrophysical signal due to an impedance mismatch at the balun feedpoint, then this re-radiated signal would be picked up by other antennas in the array and lead to excess correlation in the visibilities.
The challenge lies in formalising this mechanism and rewriting the results in terms of uncoupled quantities.

We are revisiting the derivation of their coupling model to address two issues with their results.
Their analysis implicitly assumed that the antennas were not elliptically polarized, which ultimately led to a conjugation error in their expression of the effective coupling coefficients.
Additionally, the main results of their analysis expressed the coupling coefficients in terms of antenna properties that are not commonly used in visibility simulations.
We address both of these issues in the remainder of this section by briefly re-deriving the coupled visibilities to first-order in the coupling coefficients.

\subsection{Zeroth-Order Visibilities}
\label{sec:ZEROTH-ORDER-VISIBILITIES}
The visibility $\mathbfss{V}_{ij}$ is the cross-correlation of the voltages $\mathbfit{v}_i, \mathbfit{v}_j$ measured by antennas $i$ and $j$,
\begin{align}
    \label{eq:voltage-cross-corr}
    \mathbfss{V}_{ij} = N \bigl\langle \mathbfit{v}_i \mathbfit{v}_j^\dag \bigr\rangle,
\end{align}
where $N$ is a normalization constant that converts from V$^2$ to Jy, and the angled brackets indicate an ensemble average.
For a dual-polarization instrument with feed polarizations $p,q$, the voltage vector can be written as
\begin{align}
    \label{eq:voltage-vector}
    \mathbfit{v} =
    \begin{pmatrix}
        v^p \\
        v^q
    \end{pmatrix},
\end{align}
and the polarized visibility can be written as
\begin{align}
    \label{eq:visibility-matrix}
    \mathbfss{V} =
    \begin{pmatrix}
        V^{pp} & V^{pq} \\
        V^{qp} & V^{qq}
    \end{pmatrix}.
\end{align}
The voltage measured by feed $p$ of antenna $i$ is related to the incident electric field $\mathbfit{E}$ through the antenna's `effective height' $\mathbfit{h}^p$
\begin{align}
    \label{eq:measured-voltage}
    v_i^p = \int_{4\pi} \mathbfit{h}_i^p(\hat{\mathbfit{n}}) \cdot \mathbfit{E}_i(\hat{\mathbfit{n}}) {\rm d}\Omega,
\end{align}
where $\mathbfit{E}_i(\hat{\mathbfit{n}})$ denotes the electric field coming from direction $\hat{\mathbfit{n}}$, evaluated at position $\mathbfit{x}_i$.
Since the term `effective height' is rarely used in a radio astronomy context, but the physical height of the feed above the dish is often discussed, it is important to make a distinction between the \emph{effective height} and the \emph{feed height}: the effective height characterises the electromagnetic properties of an antenna, while the feed height refers to the physical elevation of the feed above the dish centre.
\footnote{
The effective height is defined as $v = \mathbfit{h} \cdot \mathbfit{E}$ and is loosely just the ratio between the induced voltage and the incident electric field.
The nomenclature originates in the definition of the gain of a long wavelength vertical antenna (e.g., a radio tower).
More concretely, the effective height is the length of a vertical antenna that would produce the same signal as the antenna in question.
`Effective length' is sometimes also used, but here we preserve the historical usage.
}
Since visibility simulations are often performed using the antenna's peak-normalized, far-field radiation pattern $\mathbfit{A}(\hat{\mathbfit{n}})$, it is helpful to  rewrite~\autoref{eq:measured-voltage} in terms of this quantity.
This can be achieved by recognizing that the feed's effective height is proportional to its far-field radiation pattern, so we may write
\begin{align}
    \label{eq:measured-voltage-with-beam}
    v_i^p = h_{i,0}^p \int_{4\pi} \mathbfit{A}_i^p(\hat{\mathbfit{n}}) \cdot \mathbfit{E}_i(\hat{\mathbfit{n}}) {\rm d}\Omega,
\end{align}
where $h_{i,0}^p = |\mathbfit{h}_i^p(\hat{\mathbfit{n}})|_{\rm max}$ is the maximum amplitude of the effective height.
In order to make the math more tractable, we consider the case where every antenna in the array is identical and feed polarizations for an antenna are related through a simple azimuthal rotation so that $h_{i,0}^p = h_0$.

In order to obtain the zeroth-order visibilities in a familiar form, we first arrange the far-field radiation pattern into a Jones matrix
\begin{align}
    \label{eq:jones-matrix-defn}
    \mathbfss{J} =
    \begin{pmatrix}
        A^p_\theta & A^p_\phi \\
        A^q_\theta & A^q_\phi
    \end{pmatrix},
\end{align}
where $\theta,\phi$ correspond to the different electric field polarizations.
With this, the voltage measured by antenna $i$ is
\begin{align}
    \label{eq:measured-voltage-vector}
    \mathbfit{v}_i = h_0 \int_{4\pi} \mathbfss{J}(\hat{\mathbfit{n}}) \mathbfit{E}_i(\hat{\mathbfit{n}}) {\rm d}\Omega,
\end{align}
and so the visibility measured by baseline $(i,j)$ is
\begin{align}
    \label{eq:measured-visibility-without-coherence-matrix}
    \mathbfss{V}_{ij} = Nh_0^2 \int \mathbfss{J}(\hat{\mathbfit{n}}) \Bigl\langle \mathbfit{E}_i(\hat{\mathbfit{n}}) \mathbfit{E}_j(\hat{\mathbfit{n}}')^\dag \Bigr\rangle \mathbfss{J}(\hat{\mathbfit{n}}')^\dag {\rm d}\Omega {\rm d}\Omega'.
\end{align}
This can be simplified by rewriting the expectation value of the outer product of the electric field in terms of the coherency matrix and the fringe via
\begin{align}
    \label{eq:coherency-matrix}
    \bigl\langle \mathbfit{E}_i(\hat{\mathbfit{n}}) \mathbfit{E}_j(\hat{\mathbfit{n}})^\dag \bigr\rangle = \mathbfss{C}(\hat{\mathbfit{n}}) e^{-i2\pi\nu\mathbfit{b}_{ij}\cdot\hat{\mathbfit{n}}/c} \delta^{\rm D}(\hat{\mathbfit{n}}-\hat{\mathbfit{n}}'),
\end{align}
where $\mathbfit{b}_{ij} = \mathbfit{x}_j - \mathbfit{x}_i$ is the baseline separating antennas $i$ and $j$, $\nu$ is the frequency at which the visibility is measured, $c$ is the speed of light in vacuum, and $\delta^{\rm D}$ is the Dirac-delta.
Inserting~\autoref{eq:coherency-matrix} into~\autoref{eq:measured-visibility-without-coherence-matrix} casts the zeroth-order visibility into a more familiar form:
\begin{align}
    \label{eq:zeroth-order-visibility}
    \mathbfss{V}_{ij}^{(0)} = Nh_0^2 \int_{4\pi} \mathbfss{J}(\hat{\mathbfit{n}}) \mathbfss{C}(\hat{\mathbfit{n}}) \mathbfss{J}(\hat{\mathbfit{n}})^\dag e^{-i2\pi\nu\mathbfit{b}_{ij}\cdot\hat{\mathbfit{n}}/c} {\rm d}\Omega.
\end{align}

\subsection{The Scattered Field}
\label{sec:SCATTERED-FIELD}
In order to compute the first-order coupling, we must first consider how an incident electric field is re-radiated by an antenna.
An impedance mismatch between the feed and the rest of the signal chain causes part of the incident signal to be reflected, and the reflected portion is re-radiated from the feed in spherical wavefronts.
The re-radiated field $\mathbfit{E}'$ for a dipole antenna can be written as
\begin{align}
    \label{eq:scattered-field-dipole}
    \mathbfit{E}' = \frac{i\eta_0}{4\lambda} \frac{\Gamma}{R} \mathbfit{h}^* \bigl(\mathbfit{h}\cdot\mathbfit{E}\bigr) \frac{e^{ikr}}{r},
\end{align}
where $R$ is the antenna resistance, $\lambda$ is the wavelength of the radiation, $\eta_0$ is the impedance of free space, $\Gamma$ is the voltage reflection coefficient obtained from the impedance mismatch, $k = 2\pi/\lambda$ is the wavenumber, and $r$ is the distance from the scattering element.
Note that this differs from the expression in~\citet{Josaitis:2021a} (as well as other sources such as~\citealt{Sneha-2013} and~\citealt{FLOKAS}) since the effective height of an elliptically-polarized antenna in transmission is conjugated relative to its effective height in reception.
Extending this result to dual-polarization antennas, and writing $k = 2\pi\nu/c$, the  expression for the scattered electric field is
\begin{align}
    \label{eq:dual-elliptical-pol-scattered-field}
    \mathbfit{E}' = \frac{i\eta_0}{4\lambda}\frac{\Gamma}{R} \frac{e^{i2\pi\nu r/c}}{r} h_0^2 \mathbfss{J}^\dag \mathbfss{J}\mathbfit{E}.
\end{align}
Note that this has the Jones matrix appearing with its Hermitian conjugate as usually expected, and this feature was missing from the derivation provided in \citet{Josaitis:2021a}.

Taking into account that radiation from the entire sky is re-radiated, we denote the extra electric field measured at antenna $i$ due to scattering by antenna $j$ as
\begin{align}
    \label{eq:scattered-field-two-antennas}
    \mathbfit{E}'_{ij}(\hat{\mathbfit{n}}) = \xi p_{ij} (1 - \delta_{ij}) \delta^{\rm D}(\hat{\mathbfit{n}} - \hat{\mathbfit{b}}_{ij}) \mathbfss{J}(\hat{\mathbfit{b}}_{ji})^\dag \int_{4\pi} \mathbfss{J}(\hat{\mathbfit{n}}') \mathbfit{E}_j(\hat{\mathbfit{n}}') {\rm d}\Omega',
\end{align}
where $\delta_{ij}$ is the Kronecker-delta, $(1-\delta_{ij})$ is a factor forbidding self-coupling, the Dirac-delta $\delta^{\rm D}(\cdots)$ enforces that the additional electric field at antenna $i$ only comes from the direction of antenna $j$, $p_{ij} = \exp\bigl(i2\pi\nu \tau_{ij}\bigr)/b_{ij}$ is a spherical wave propagation term, $\tau_{ij} = b_{ij}/c$ is the delay corresponding to the time it takes the scattered radiation to propagate from antenna $j$ to antenna $i$, and $\xi$ bundles together additional factors that are common to all antennas.
For simplicity, we assume that each antenna has an unobstructed view of every other antenna in the array, so the total excess electric field at any antenna can be obtained by summing~\autoref{eq:scattered-field-two-antennas} over the entire array, giving the total electric field at any antenna $i$ as
\begin{align}
    \label{eq:total-electric-field}
    \mathbfit{E}^{\rm tot}_i(\hat{\mathbfit{n}}) = \mathbfit{E}_i(\hat{\mathbfit{n}}) + \sum_k \mathbfit{E}'_{ik}(\hat{\mathbfit{n}}).
\end{align}

\subsection{First-Order Coupled Visibilities}
\label{sec:COUPLED-VISIBILITIES}
We compute the effect of re-radiation coupling by first inserting~\autoref{eq:total-electric-field} into~\autoref{eq:measured-visibility-without-coherence-matrix}.
Keeping only terms to first-order in the scattered field, this gives
\begin{align}
    \nonumber
    \mathbfss{V}_{ij}^{(1)} &= Nh_0^2 \int \mathbfss{J}(\hat{\mathbfit{n}}) \Bigl\langle \mathbfit{E}_i(\hat{\mathbfit{n}}) \mathbfit{E}_j(\hat{\mathbfit{n}}')^\dag \Bigr\rangle \mathbfss{J}(\hat{\mathbfit{n}}')^\dag {\rm d}\Omega {\rm d}\Omega' \\
    \nonumber
    &+ Nh_0^2 \sum_k \int \mathbfss{J}(\hat{\mathbfit{n}}) \Bigl\langle \mathbfit{E}'_{ik}(\hat{\mathbfit{n}}) \mathbfit{E}_j(\hat{\mathbfit{n}}')^\dag \Bigr\rangle \mathbfss{J}(\hat{\mathbfit{n}}')^\dag {\rm d}\Omega {\rm d}\Omega' \\
    &+ Nh_0^2 \sum_k \int \mathbfss{J}(\hat{\mathbfit{n}}) \Bigl\langle \mathbfit{E}_i(\hat{\mathbfit{n}}) \mathbfit{E}'_{jk}(\hat{\mathbfit{n}}')^\dag \Bigr\rangle \mathbfss{J}(\hat{\mathbfit{n}}')^\dag {\rm d}\Omega {\rm d}\Omega'.
    \label{eq:first-order-vis-unevaluated}
\end{align}
The first term in this expression is just the zeroth-order visibility $\mathbfss{V}_{ij}^{(0)}$.
Using~\autoref{eq:coherency-matrix} and the Dirac-delta in~\autoref{eq:scattered-field-two-antennas} allows us to rewrite the integral in the second term as
\begin{align}
    \nonumber
    \int &\mathbfss{J}(\hat{\mathbfit{n}}) \Bigl\langle \mathbfit{E}'_{ik}(\hat{\mathbfit{n}}) \mathbfit{E}_j(\hat{\mathbfit{n}}')^\dag \Bigr\rangle \mathbfss{J}(\hat{\mathbfit{n}}')^\dag {\rm d}\Omega {\rm d}\Omega' \\
    \nonumber
    &= \xi p_{ik} (1-\delta_{ik}) \mathbfss{J}(\hat{\mathbfit{b}}_{ik}) \mathbfss{J}(\hat{\mathbfit{b}}_{ki})^\dag \\
    &\times \int_{4\pi} \mathbfss{J}(\hat{\mathbfit{n}}) \mathbfss{C}(\hat{\mathbfit{n}}) \mathbfss{J}(\hat{\mathbfit{n}})^\dag e^{-i2\pi\nu\mathbfit{b}_{kj}\cdot\hat{\mathbfit{n}}/c} {\rm d}\Omega.
\end{align}
The integral in this expression is just the visibility for baseline $(k,j)$, so the above expression can be written more compactly as
\begin{align}
\label{eq:coupled_terms}
    \int \mathbfss{J}(\hat{\mathbfit{n}}) \Bigl\langle \mathbfit{E}'_{ik}(\hat{\mathbfit{n}}) \mathbfit{E}_j(\hat{\mathbfit{n}}')^\dag \Bigr\rangle \mathbfss{J}(\hat{\mathbfit{n}}')^\dag {\rm d}\Omega {\rm d}\Omega' = \mathbfss{X}_{ik} \mathbfss{V}_{kj}^{(0)},
\end{align}
where the coupling coefficient $\mathbfss{X}_{ik}$ is
\begin{align}
    \label{eq:coupling-coeff-before-rewrite}
    \mathbfss{X}_{ik} = \frac{i\eta_0}{4\lambda} \frac{\Gamma}{R} \frac{e^{i2\pi\nu\tau_{ik}}}{b_{ik}} h_0^2 \mathbfss{J}\bigl(\hat{\mathbfit{b}}_{ik}\bigr) \mathbfss{J}\bigl(\hat{\mathbfit{b}}_{ki}\bigr)^\dag.
\end{align}
We get a similar result for the integral in the third term in~\autoref{eq:first-order-vis-unevaluated}:
\begin{align}
\label{eq:coupled_terms_dag}
    \int \mathbfss{J}(\hat{\mathbfit{n}}) \Bigl\langle \mathbfit{E}_i(\hat{\mathbfit{n}}) \mathbfit{E}'_{jk}(\hat{\mathbfit{n}}')^\dag \Bigr\rangle \mathbfss{J}(\hat{\mathbfit{n}}')^\dag {\rm d}\Omega {\rm d}\Omega' = \mathbfss{V}_{ik}^{(0)} \mathbfss{X}_{jk}^\dag,
\end{align}
and so the first-order visibilities can be expressed via
\begin{align}
    \label{eq:first-order-visibility}
    \mathbfss{V}_{ij}^{(1)} = \mathbfss{V}_{ij}^{(0)} + \sum_k\Bigl( \mathbfss{V}_{ik}^{(0)} \mathbfss{X}_{jk}^\dag + \mathbfss{X}_{ik} \mathbfss{V}_{kj}^{(0)}\Bigr).
\end{align}

\subsection{Coupling Coefficients}
\label{sec:COUPLING-COEFFICIENTS}
Visibility simulators often do not have access to the effective height, but rather the far-field radiation pattern, so it is helpful to rewrite $h_0$ in terms of something that can be computed from the radiation pattern.
The effective height $\mathbfit{h}$ is related to the current $I$ in the antenna and the radiation pattern $\mathbfit{F}$ produced by the current via $I\mathbfit{h} = \mathbfit{F}$, and so $Ih_0 = F_0$.
The current is related to the power lost due to resistive losses in the antenna 
\begin{align}
    \label{eq:radiated-power}
    P_{\rm T} = \frac{1}{2} |I|^2 R.
\end{align}
Assuming that all of this power is radiated away by the antenna, we can also express it in terms of the radiation intensity $U$ integrated over the full sky
\begin{align}
    \label{eq:power-integral}
    P_{\rm T} = \int_{4\pi} U {\rm d}\Omega = \frac{\eta_0 F_0^2}{8\lambda^2} \int_{4\pi} \frac{F^2}{F_0^2} {\rm d}\Omega = \frac{\eta_0 F_0^2}{8\lambda^2} \Omega_{\rm p},
\end{align}
where the second equality comes from $8\lambda^2 U = \eta_0 |\mathbfit{F}|^2$ and $\Omega_{\rm p}$ is the beam area.
Using $Ih_0 = F_0$ with the above two equations gives the following relation
\begin{align}
\label{eq:coeff-sub}
    \frac{\eta_0 h_0^2}{4\lambda^2 R} = \frac{1}{\Omega_{\rm p}},
\end{align}
which we can use to rewrite the coupling coefficients as
\begin{align}
    \label{eq:coupling-coefficients}
    \mathbfss{X}_{ik} = \frac{i\Gamma}{\Omega_{\rm p}} \frac{e^{i2\pi\nu\tau_{ik}}}{u_{ik}} \mathbfss{J}\bigl(\hat{\mathbfit{b}}_{ik}\bigr) \mathbfss{J}\bigl(\hat{\mathbfit{b}}_{ki}\bigr)^\dag,
\end{align}
where $u_{ik} = b_{ik}/\lambda$ is the baseline length in units of wavelengths.
Using this expression, only the impedance and far-field radiation pattern of an isolated antenna need to be exported from an electromagnetic simulator, since each term in the above expression can be computed from these quantities or metadata about the interferometric array.

\subsection{Simulation of Mutual Coupling}
\label{sec:SIMULATION}
While the first-order coupling model provides a computationally tractable way to simulate mutual coupling for large arrays, a direct implementation of~\autoref{eq:first-order-visibility} and~\autoref{eq:coupling-coefficients} can require fairly long compute times.
There are a few symmetries in the design of the HERA array and the individual antennas that can be leveraged to greatly reduce the computational complexity of a coupling simulation, which we explain in the following paragraphs.

The first approximation we can make appeals to the symmetry of the HERA antenna.
Since the HERA dish is a circular paraboloid and the feed is symmetric under 90$^\circ$ rotations about its vertical axis, the HERA antenna is symmetric under 180$^\circ$ rotations about boresight.
Consequently, rotating the HERA antenna in this way must leave the antenna's radiation pattern unchanged.
Decomposing the antenna radiation pattern into its vector components via $\mathbfit{F} = F_\theta \hat{\boldsymbol{\theta}} + F_\phi \hat{\boldsymbol{\phi}}$, this implies
\begin{equation}
    F_\theta(\theta,\phi) \hat{\boldsymbol{\theta}} + F_\phi(\theta,\phi) \hat{\boldsymbol{\phi}} = F_\theta(\theta,\phi+\pi) \hat{\boldsymbol{\theta}} - F_\phi(\theta,\phi+\pi) \hat{\boldsymbol{\phi}},
\end{equation}
since this type of transformation leaves $\hat{\boldsymbol{\theta}}$ unchanged but sends $\hat{\boldsymbol{\phi}} \rightarrow -\hat{\boldsymbol{\phi}}$.
Now, restricting our attention to the plane of the horizon, if we approximate the feed as a horizontal dipole, then the radiation field is purely azimuthal and so it follows that $\mathbfit{F}(\pi/2,\phi) = -\mathbfit{F}(\pi/2,\phi+\pi)$.
While this is not an entirely accurate description of the HERA antenna's radiation pattern, it turns out to be a reasonable approximation and provides the useful result
\begin{equation}
    \mathbfss{J}(\hat{\mathbfit{b}}_{ij}) \approx -\mathbfss{J}(\hat{\mathbfit{b}}_{ji}).
\end{equation}
Under this assumption, the coupling coefficients become symmetric under exchanging indices, so that
\begin{align}
    \label{eq:coupling-coeff-symmetry}
    \mathbfss{X}_{ik} = \mathbfss{X}_{ki}.
\end{align}
Using this and indexing the visibility and coupling matrices by antenna-polarization pairs, where even and odd indices correspond to different feed polarisations, allows us to write the first-order visibilities as a simple matrix product:
\begin{equation}
    \label{eq:simulation-equation}
    \mathbfss{V}^{(1)} = \mathbfss{V}^{(0)} + \mathbfss{V}^{(0)}\mathbfss{X}^\dag + \Bigl(\mathbfss{V}^{(0)}\mathbfss{X}^\dag\Bigr)^\dag.
\end{equation}
Given the coupling matrix and the zeroth-order visibilities, the first-order visibilities can thus be efficiently computed with a matrix product, and employing the symmetry approximation provides us with a factor of two speedup.

Computing the coupling matrix can be computationally expensive, since it is typically expensive to interpolate the beam and the coupling expression suggests that this interpolation would need to be performed once for each baseline in the array.
Redundant array configurations like HERA, however, have many fewer unique baseline orientations than total baselines, and this can be exploited to reduce computational demands.
To take advantage of this, we only interpolate the beam once for each unique baseline orientation in the array, temporarily cache the result, and retrieve the cached value when computing the elements of the coupling matrix.
Since the HERA array is roughly planar, the interpolation routine used in computing the coupling coefficients is further sped up by using a truncated version of the beam which has full azimuth information but only a few pixels above and below the horizon.
\footnote{
    A few pixels above and below the horizon are needed for handling baselines with a small vertical component.
}
This interpolation scheme produces the same values as interpolating with the full sky, but greatly reduces the computational cost of constructing and evaluating the interpolation spline.
All other quantities only need to be computed once or are fast to compute on-the-fly.
This algorithm is implemented in a module of the HERA instrument simulator, \textsc{hera\_sim}.\footnote{\url{https://github.com/HERA-Team/hera_sim}}

\section{HERA Phase II Observations and Simulations}
\label{sec:OBSERVATIONS}

\begin{figure}
    \centering
    \includegraphics[width=\columnwidth]{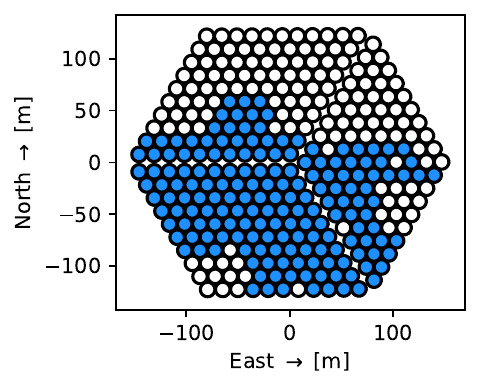}
    \caption{
        Array layout for the 320 core HERA antennas, shown in a local coordinate system relative on the array's centre.
        Antennas highlighted in blue were used in the analysis presented in this paper.
    }
    \label{fig:Array_Plot}
\end{figure}

HERA is a 350-element interferometer, built in South Africa’s Karoo Radio Astronomy Reserve and co-located at the site of the high band portion of the Square Kilometer Array.
Each element of the interferometer consists of a 14-m parabolic dish outfitted with a Vivaldi feed~\citep{Fagnoni_2021}, observing over 50--250 MHz. 
The hexagonal core contains 320 antennas.
The remaining 30 antennas are outriggers are spaced at intervals of approximately 140 m away from each other and away from the core, arranged in two concentric hexagonal rings.

\subsection{Data Description}
\label{sec:DATA_DESCRIPTION}
We use 12 nights of data taken during the sixth HERA observing season (2022--23). 
This dataset contains 183 antennas, 175 of which are located in the core of the array.
\autoref{fig:Array_Plot} shows the HERA core in local coordinates; the core antennas present in this dataset are highlighted in blue.
As in \citet{Josaitis:2021a}, we limit our analysis in this paper to 145--165 MHz.
The predicted emission signature of the 21-cm brightness temperature signal during reionisation falls within this frequency range.
\autoref{table:obs_params} summarizes various observational and data parameters pertaining to our analysis.

Here, we provide a brief overview of the HERA data processing pipeline for antennas in the core of the array; for a full description of the analysis pipeline, see HERA Memo 124~\citep{IDR21_Memo}\footnote{\url{https://reionization.org/manual_uploads/HERA124_H6C_IDR_2_Memo_v3.pdf}}.
The processed data was used to produce the comparisons between the data and simulations shown in~\autoref{fig:compare_data_sim_fornax} and~\autoref{fig:compare_data_sim_gc}, which we discuss in more detail in Section~\ref{sec:COMPARISON}.
First, RFI is excised using a flagging algorithm which looks for outliers in array-averaged autocorrelations.
Next, the flagged data is calibrated using a combination of redundant baseline calibration~\citep{2020MNRAS.499.5840D} and a constrained absolute calibration whereby the overall flux scale and phase centre is set using visibility simulations.
The per-antenna complex gains produced by the calibration pipeline are then smoothed across time and frequency to avoid introducing additional structure into the data.
A second RFI-flagging step, which uses delay filtering to search for broadband RFI, is performed on the calibrated data.
The flagged and calibrated data are then binned in Local Sidereal Time (LST) and averaged over nights. 
In order to remove any remaining discontinuities from missing data, inpainting is then performed on the LST-averaged data to fill in times and frequencies containing no data (Chen et al. in preparation).
The inpainting is performed with a technique that uses unflagged data to estimate the data covariance, which is then used to `interpolate' over gaps in the data (see e.g., Section 3.4 of~\citealt{Pagano_2023} for details).

The data in~\autoref{fig:autos_comparison}, which compares the autocorrelation for an outrigger antenna against that of a core antenna, were treated slightly differently, since the outriggers were still undergoing active commissioning and had not been calibrated.
For the outrigger autocorrelations, we instead performed inpainting over flagged channels on a per-night basis, then peak-normalized and stacked the data over nights.
This calibration is not sufficient for an EoR power spectrum analysis, but suffices for a qualitative examination of mutual coupling features in the outrigger data in the absence of proper calibration solutions.


\subsection{Simulation Description}
\label{sec:SIMULATION-DESCRIPTION}
To perform the data-to-model comparison in Section~\ref{sec:COMPARISON}, we simulate first-order mutual coupling for the 175 core antennas which were present in the data. 
The zeroth-order visibility simulations use a perfectly redundant array layout corresponding to the 175 core antennas in the data.
For the beam model, we use the far-field radiation pattern from a CST simulation of the Phase II HERA antenna~\citep{Fagnoni_2021}.
The simulations use a foreground-only sky that is a hybrid synthetic-empirical model.
One component of the sky model consists of point sources uniformly distributed on the sky with source fluxes randomly assigned between 1 mJy and 100 Jy in a manner consistent with source counts reported in~\citet{Franzen_2019}; sources below 1 Jy are binned into a HEALPix~\citep{Gorski_2005} map and are used as an additional `confusion noise' component in the diffuse sky model.
In addition to the synthetic point source catalog, we include a simplified model of bright radio sources which are individually treated as point sources with fluxes, positions, and spectral indices as reported in Table 2 of~\citet{Hurley-Walker_2017}.
The diffuse component of the sky model uses the 2008 Global Sky Model~\citep{Oliveira-Costa_2008} in addition to the confusion noise component leftover from the synthetic source model.
Visibility simulation was then performed with \textsc{matvis}~\citep{Kittiwisit_2023} using tools provided in \textsc{hera\_sim}.
The simulations computed visibilities for a full sidereal day sampled at a 5-second cadence over the extended HERA bandwidth, but only data from the 145--165 MHz subband and 0--4.9 hr LST range was used for this work.

First-order coupling was applied to the zeroth-order simulations according to the prescription described in Section~\ref{sec:FORMALISM}, using the tools available in \textsc{hera\_sim}.
Note that the coupling coefficients in the first-order model require knowledge of the voltage reflection spectrum $\Gamma$ as well as the beam integrals $\Omega_{\rm p}$.
The reflection spectrum was calculated from the antenna impedance $Z_{\rm ant}$ and receiver impedance $Z_{\rm rec}$ reported in~\citet{Fagnoni_2021} via
\begin{equation}
    \Gamma = \frac{Z_{\rm rec} - Z_{\rm ant}}{Z_{\rm rec} + Z_{\rm ant}}.
\end{equation}
The beam integrals were computed using the same beam model that was used for the zeroth-order visibility simulations.
The horizon beam terms $\mathbfss{J}(\hat{\mathbfit{b}}_{ij})$ were obtained by interpolating a truncated version of the beam used in the zeroth-order visibility simulations, as described in Section~\ref{sec:FORMALISM}.

In addition to foreground and coupling simulations, we simulated visibilities for a mock cosmological 21-cm signal to be used as a comparison point when investigating the effects of mutual coupling and fringe-rate filtering on power spectrum estimates in Section~\ref{sec:POWER-SPECTRUM}.
For the mock 21-cm signal, we used a realization from a Gaussian random field with a power-law power spectrum that is consistent with the power spectrum predicted by~\citet{Munoz:2022}.
All of the instrument parameters for the simulated 21-cm visibilities are the same as those used for the foreground simulations.
Importantly, the simulated 21-cm visibilities were not included in the uncoupled or coupled visibilities; the mock 21-cm signal is simply used as a device for understanding the relative strength of the predicted coupling features and a realistic cosmological 21-cm signal.

\begin{table}
\caption{The HERA Phase II observation parameters used in this analysis}
\centering
\begin{tabular}{ |p{3.5cm}||p{4cm}| }

 Parameter & Value \\
 \hline
 Telescope Latitude & -30$^\circ$ 43' 17" \\
 Telescope Longitude & 21$^\circ$ 25' 41" \\
 Number of Antennas & 183 \\
 Number of Nights & 12 \\
 Frequency Range & 145--165 MHz \\
 LST Ranges & 23.9--2.4 hr; 2.4--4.9 hr\\
 Channel Width (RBW)  & 122.07 kHz \\
 Integration Time & 10.7 s \\
 Dish Diameter & 14 m \\
 Feed Type & Vivaldi \\
 Instrument polarization & North-South \\
\hline
\end{tabular}
\label{table:obs_params}
\end{table}

\section{Evidence of First-Order Mutual Coupling in HERA Data}
\label{sec:EVIDENCE}
While~\citet{Josaitis:2021a} provided a detailed study of how first-order coupling manifests in simulated visibilities, first-order coupling has not yet been compared against systematic features seen in observed data.
In this section, we explore whether the systematic features predicted by the first-order coupling model accurately capture the systematic features seen in the data.
While visibilities are naturally measured as a function of frequency and time, we opt to perform our analysis in fringe-rate versus delay space, since it was shown in~\citet{Kern:2019} and~\citet{Kern2020a} that this space naturally separates various systematic features.
Since fringe-rate is the Fourier dual to time and delay is the Fourier dual to frequency, visibilities in fringe-rate versus delay space may be obtained from a more traditional time versus frequency `waterfall' by performing a (tapered) two-dimensional Fourier transform via\footnote{A more pedagogical review of transforming to fringe-rate versus delay space is provided in Appendix~\ref{sec:FRATE-DELAY-TRANSFORM}.}
\begin{equation}
    \label{eq:fr-dly-transform}
    \tilde{V}_{ij}(\tau,f_r) = \int T(\nu) W(t) V_{ij}(\nu, t) e^{-i2\pi\nu\tau} e^{-i2\pi f_r t} {\rm d}\nu {\rm d}t,
\end{equation}
where $V_{ij}(\nu,t)$ is the visibility at frequency $\nu$ and time $t$, $\tau$ is delay, $f_r$ is fringe-rate, $T(\nu)$ is the taper used for the delay transform, and $W(t)$ is the taper used for the fringe-rate transform.

\citet{Josaitis:2021a} similarly performed the bulk of their analysis in fringe-rate versus delay space and identified two key features associated with first-order mutual coupling in this space.
The first is a cross feature arising from fringing emission, in which we observe excess power at non-zero fringe-rates, manifesting as an `X' or diagonal `slash' shape in fringe-rate versus delay space.
The second mutual coupling feature is a `bar' feature with excess power near zero fringe-rate across a broad range of delays, associated with non- or slowly-fringing emission.
These extended cross and bar features cannot come from the sky, since the uncoupled signal is dominated by spectrally smooth foreground emission rotating with the sky, which limits the extent of the signal in both fringe-rate and delay.
We see evidence of both the bar and the cross features in the data, and thus conclude that mutual coupling is at least partially responsible for some of the systematic excesses seen in the data. 
The extent to which we are able to attribute the observed excess power to mutual coupling, however, is somewhat uncertain.
Excess power near zero fringe-rate may be sourced from other time-stable broadband features, such as crosstalk between cables or a non-smooth bandpass.
Additionally, as discussed further in Section~\ref{sec:AMPLITUDE}, the cross feature is somewhat brighter in the data than predicted by the re-radiation model, so there may be additional mechanisms at play.

\subsection{Data-to-Model Comparison}
\label{sec:COMPARISON}

\begin{figure*}
    \centering
    \includegraphics[width=1.0\textwidth]{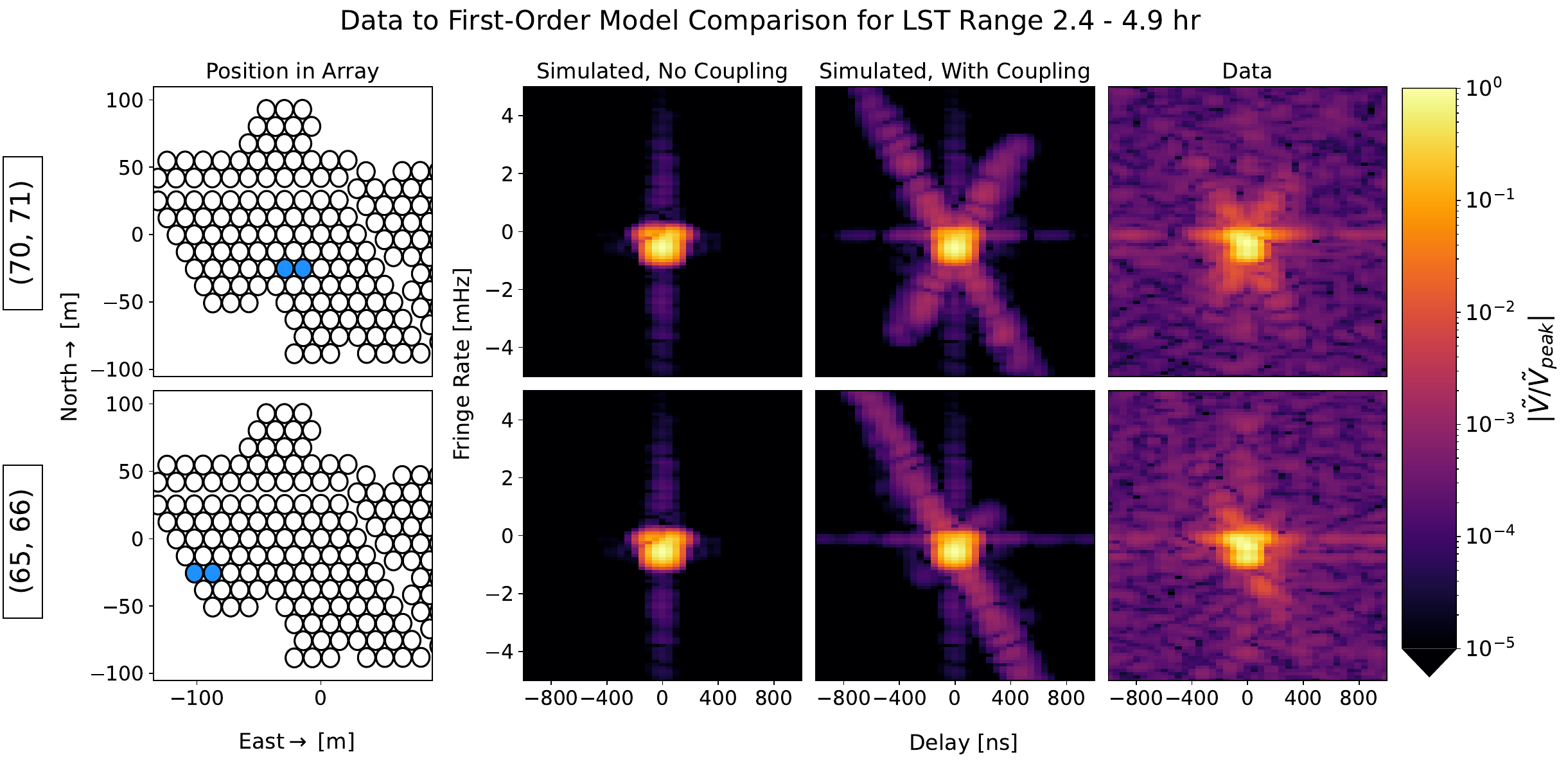}
    \caption{
        A comparison of simulated first-order coupling and HERA Phase II Data for two baselines.
        From left-to-right: the first column shows the array layout, with the antennas used to form the baselines highlighted in blue; the second column shows the simulated zeroth-order (or uncoupled) visibilities; the third column shows the simulated first-order coupled visibilities; the fourth column shows the observed data.
        Both rows show visibilities for a 14.6-m baseline; the baseline in the top row is formed with antennas in the centre of the array, while the baseline in the bottom row is formed with antennas at the edge of the array.
        The data are taken during the LST range 2.4--4.9 hr, when Fornax A is visible to the array.
        We observe the cross feature predicted by the model in the data, which is not predicted by the zeroth-order visibility simulation.
        The amplitude of the data and simulation agree to within roughly an order of magnitude.
        We observe a mostly symmetrical cross feature in baseline (70, 71) (top row) due to its position near the centre of the array, since this baseline sees coupling from antennas in all directions.
        By contrast, the asymmetrical shape of the arms of the cross for baseline (94, 112) (bottom row) arises from its position near the edge of the array, as most of the coupled antennas are to the east of the baseline.
        \autoref{fig:plot_with_dots} demonstrates this effect in greater detail.
    }
    \label{fig:compare_data_sim_fornax}
\end{figure*}

\begin{figure*}
    \centering
    \includegraphics[width=1.0\textwidth]{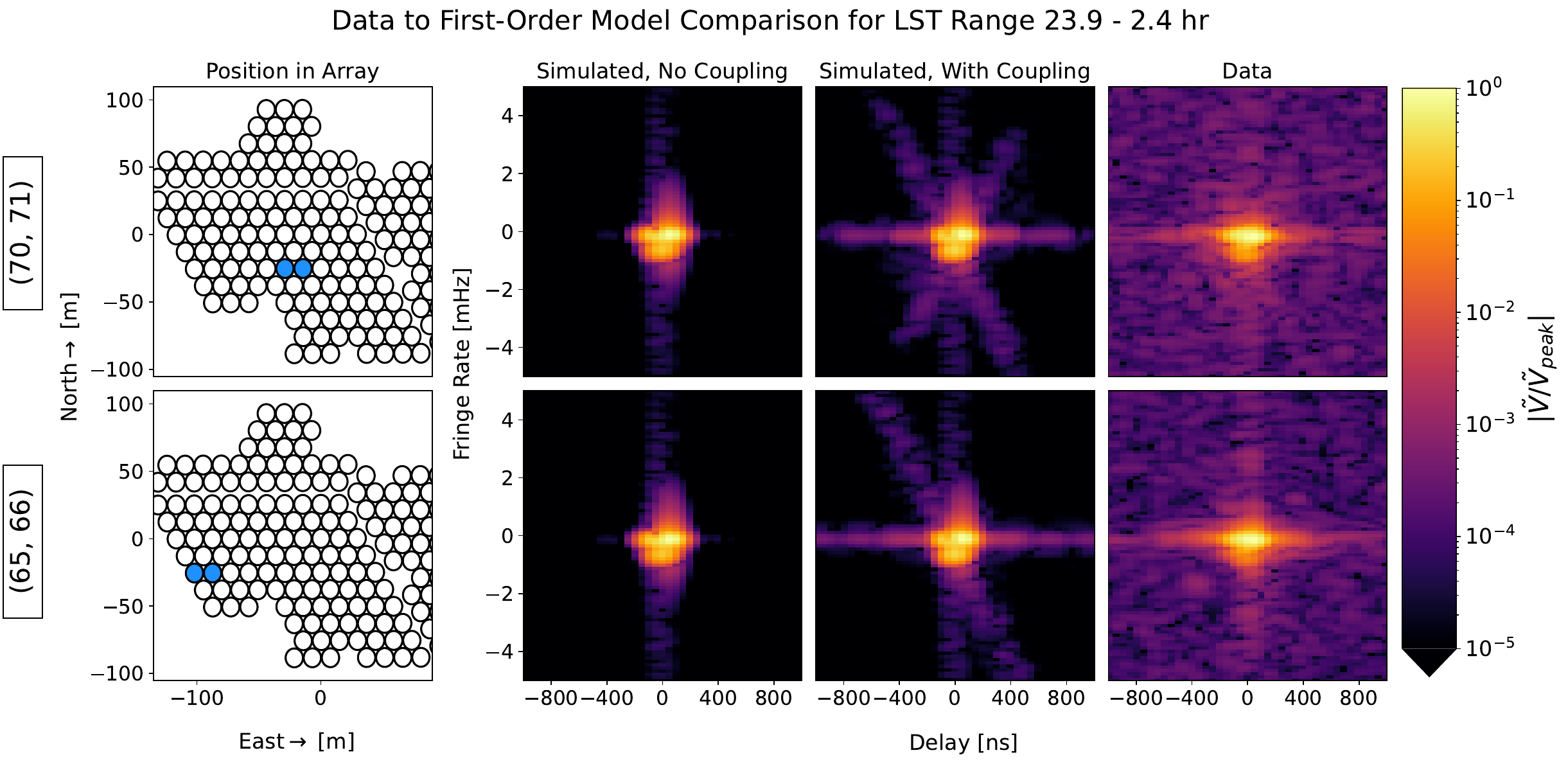}
    \caption{
        A comparison of simulated first-order coupling and HERA Phase II Data for the same baselines as in~\autoref{fig:compare_data_sim_fornax}.
        The data are taken during the LST range 23.9--2.4 hr, when the galactic centre is on the horizon.
        In this figure, we see the bar feature at zero fringe-rate more prominently.
    }
    \label{fig:compare_data_sim_gc}
\end{figure*}

\autoref{fig:compare_data_sim_fornax} and~\autoref{fig:compare_data_sim_gc} compare HERA Phase II data against noiseless first-order coupling simulations between 145--165 MHz in two LST ranges, with one spanning 23.9--2.4 hr and the other spanning 2.4--4.9 hr. 
During the 2.4--4.9 hr observation, Fornax A passes through HERA's field of view; though Fornax A only passes through the sidelobes of the primary beam, it is still bright enough to dominate the telescope response.
During the 23.9--2.4 hr observation, the galactic centre is setting on the horizon, which produces a strong `pitchfork effect' in the uncoupled visibilities~\citep{WEDGE_NITHYA_2015}. 
These observations demonstrate the cross and bar feature respectively, which we explore in more detail in Section~\ref{sec:PREDICTIONS}.

We observe good qualitative agreement between the data and first-order coupling simulation: the shape of the coupling features match, and we see excess power in the same regions of fringe-rate versus delay space.  
The amplitude of these features agree to within roughly an order of magnitude, but it is clear that the coupling amplitude is higher in the data than is predicted by the simulations.
We explore the consequences of these observations further in Section~\ref{sec:AMPLITUDE}.

\begin{figure}
    \centering
    \includegraphics[width=\columnwidth]{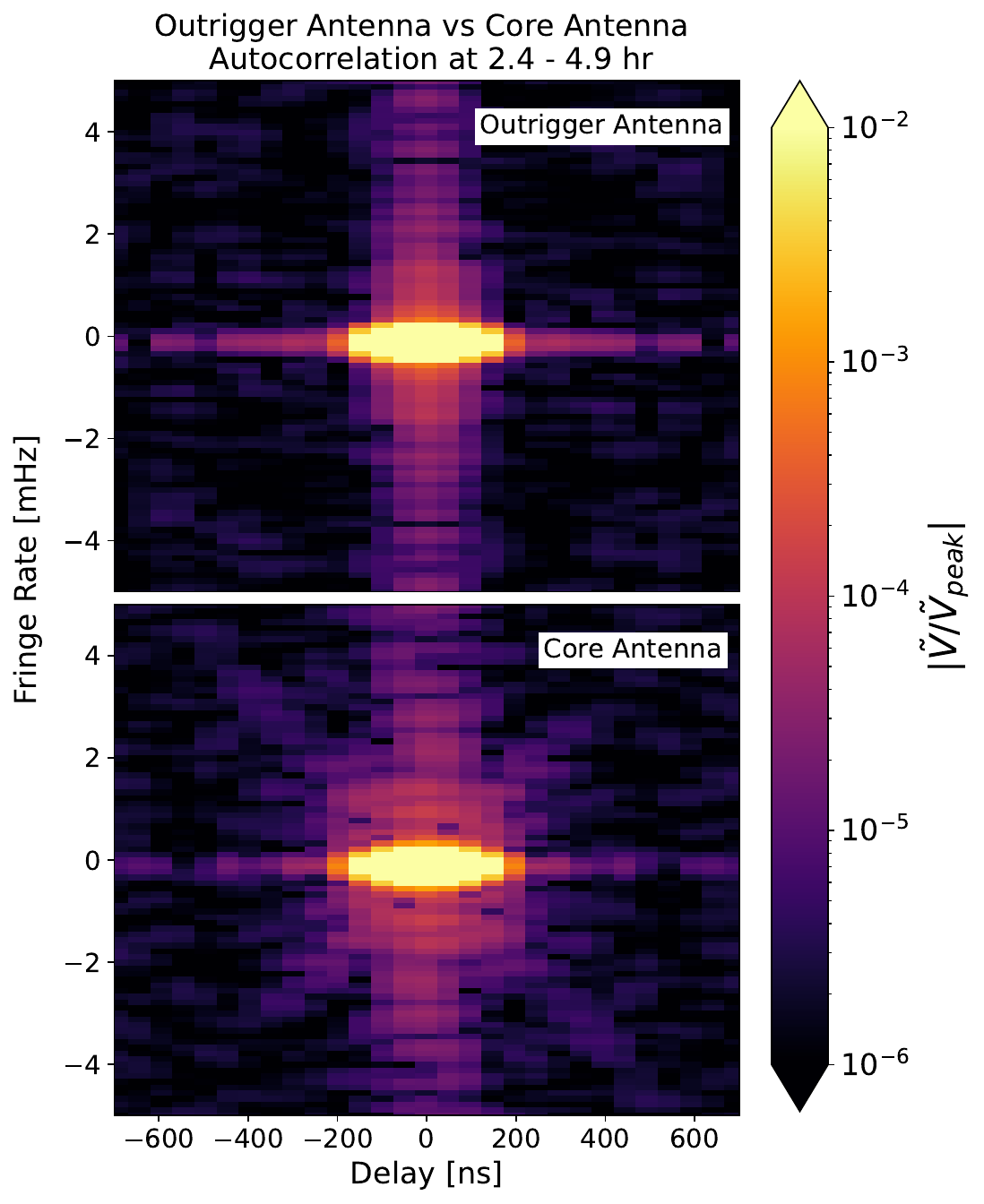}
    \caption{
        A comparison of autocorrelations from an outrigger (left) and core antenna (right) for LST range 2.4--4.9 hr, at 145--165 MHz.
        The outrigger's nearest neighboring antenna is ~150 meters away.
        The cross predicted in \citet{Josaitis:2021a}, and shown in both the data and model in~\autoref{fig:compare_data_sim_fornax}, is clearly absent in the outrigger autocorrelation.
    }
    \label{fig:autos_comparison}
\end{figure}

To confirm whether the features we observe are indeed attributable to mutual coupling, we compare data from an outrigger antenna to data from an antenna at the centre of the array.
The outrigger antenna's nearest neighbor is over 140 m away, so it can be taken as an approximate `no coupling' case. 
For the Fornax-dominated LST range (2.4--4.9 hr), we predict that the outrigger autocorrelation should not have the cross feature, but this feature should be present in the autocorrelation of a core antenna.
\autoref{fig:autos_comparison} demonstrates that this is indeed the case, which suggests that the extra structure seen in~\autoref{fig:compare_data_sim_fornax} and~\autoref{fig:compare_data_sim_gc} may be attributed to mutual coupling.
Note, however, that there is still a bar feature seen in the outrigger autocorrelation; it is likely that at least a good fraction of this is not due to mutual coupling, but instead is a consequence of the processing applied to the outrigger autocorrelations (described in Section~\ref{sec:DATA_DESCRIPTION}) not ensuring that these autocorrelations were spectrally smooth.

\subsection{The Structure of Coupling in Delay and Fringe Rate Space}
\label{sec:PREDICTIONS}

Both the cross and the bar feature can be understood as copies of the zeroth-order visibilities of other baselines coupling in at delays corresponding to the light travel time between the coupled antennas. 
Given only the array geometry, we can use this fact to predict which regions of fringe-rate versus delay space will have excess power due to mutual coupling.

We begin by revisiting the equations for the coupled visibilities (\ref{eq:first-order-visibility}) and coupling coefficients (\ref{eq:coupling-coefficients}).
Combining these equations, we can write the coupled visibilities as
\begin{equation}
    \label{eq:coupled_vis_rewrite}
    \mathbfss{V}_{ij}^{(1)} = \mathbfss{V}_{ij}^{(0)} + \sum_k\Bigl( \mathbfss{V}_{ik}^{(0)} \mathbfss{A}_{jk}^\dag e^{-i2\pi\nu\tau_{jk}} + \mathbfss{A}_{ik} \mathbfss{V}_{kj}^{(0)} e^{i2\pi\nu\tau_{ik}}\Bigr),
\end{equation}
where the $\mathbfss{A}_{ij}$ terms bundle together the antenna-related terms and the $r^{-1}$ decay factor associated with the assumed spherical wave propagation of the re-radiated field.
When taking the delay transform of~\autoref{eq:coupled_vis_rewrite}, the exponential terms $e^{\pm i2 \pi \nu \tau_{ik}}$ cause a Fourier shift of the power from the coupled visibilities along the delay axis to $\tau = \pm\tau_{ik}$.
This is equal to the time of flight of the coupled radiation from antenna $k$ to antenna $i$, $\tau_{ik} = b_{ik}/c$, and thus we only need to know the array geometry to predict which delay modes will have extra power added to them through mutual coupling.
Note that in addition to the Fourier shift, the convolution theorem implies that the delay structure of the zeroth-order visibilities that are coupled in through re-radiation will be broadened somewhat due to the spectral structure in $\mathbfss{A}$.

Along the fringe-rate axis, visibilities corresponding to baselines of different lengths and orientations will have power concentrated in different regions of fringe-rate space.
To understand why, first recall that fringe-rate is the Fourier dual of the time axis and consider the time-dependent behavior of sources on the sky.
Since HERA is a drift-scan telescope, point sources appear to move from East to West at the angular velocity of the Earth's rotation in the telescope frame of reference. 
\citet{Parsons_2016} demonstrated that for a source located at $\hat{\mathbfit{n}}$, a baseline $\mathbfit{b}$ will measure an `instantaneous fringe-rate'
\begin{equation}
\label{eq:fringe_rate}
    f = \frac{\nu}{c} \hat{\mathbfit{n}} \cdot (\mathbfit{b} \times \boldsymbol{\omega}_\oplus)
\end{equation}
where $\nu$ is the frequency of the measurement and $ \boldsymbol{\omega}_\oplus$ is the angular velocity vector of the Earth's rotation. 
For HERA, the antenna sensitivity peaks at zenith, so, assuming that there is no off-zenith source which dominates the telescope response, the visibility's fringe-rate response will peak at 
\begin{equation}
\label{eq:f_peak}
    f_{\rm peak} = -\frac{\omega_\oplus \nu}{c} \cos(\delta) b_{\rm EW}
\end{equation}
where $\delta$ is the latitude of the telescope and $b_{\rm EW}$ is the (signed) East-West projection of the baseline, with $b_{\rm EW}$ positive for baselines that point eastward.

For the zeroth-order visibility $\mathbfss{V}_{kj}^{(0)}$, which couples into $\mathbfss{V}_{ij}^{(1)}$ with the unconjugated coupling coefficient $\mathbfss{X}_{ik}$, the coupled power will be concentrated around the point 
\begin{equation}
\label{eq:tau_f_points}
    (\tau_{\rm peak}, f_{\rm peak}) = \left( \frac{b_{ik}}{c}, -\frac{\omega_\oplus \nu}{c} \cos(\delta)  b_{kj,{\rm EW}} \right).
\end{equation}
This can be used to predict the shape and extent of the cross coupling feature, and requires only knowledge of the array geometry and observatory latitude.
A worked example for a simple scenario is provided in Appendix~\ref{sec:worked_example}.

\autoref{fig:plot_with_dots} plots the predicted $(\tau_{\rm peak}, f_{\rm peak})$ points using~\autoref{eq:tau_f_points} for one baseline during the observation spanning 2.4--4.9 hr LST.
We find that there is excess power associated with many of the grey dots plotted in~\autoref{fig:plot_with_dots}, though the grey dots do not exactly coincide with all regions that contain excess power.
This may be due to the fact that the structure of the uncoupled visibilities is not confined to a single point in fringe-rate versus delay space, but rather has some spread, or it may be indicative of additional mechanisms producing the excess power.
Additionally, the grey dots extend deep into the noise-dominated regions, suggesting that there are additional coupling features that will be uncovered with more sensitive data.

The bar-like coupling feature is more prominent when the uncoupled visibilities are dominated by emission from sources at regions of zero (or nearly zero) instantaneous fringe-rate.
For HERA, this would come from either the South Celestial Pole, which is faint in the radio~\citep{Sironi:1991}, or from the horizon; since $\hat{\mathbfit{n}}$ and $\mathbfit{b}$ are coplanar along the horizon,~\autoref{eq:fringe_rate} predicts that many regions along the horizon contribute to fringe-rate modes near zero.
Since the galactic plane is incredibly bright and sets around LST $\sim$ 0 hr, the visibilities for the LST 23.9--2.4 hr field contain a large amount of power near zero fringe-rate, despite the greatly suppressed antenna response at the horizon.
As a result, this increased power near zero-fringe rate is summed together to form the bar feature observed in~\autoref{fig:compare_data_sim_gc}. 

Note that although the coupling mechanism is the same for the data presented in~\autoref{fig:compare_data_sim_fornax} and~\autoref{fig:compare_data_sim_gc}, the morphology of the coupling features depends both on where the antennas are located in the array, as well as how the sky intensity is distributed over the duration of the observation.
In~\autoref{fig:compare_data_sim_fornax}, although we see a prominent cross feature in both baselines, the symmetry of the cross feature differs between the two baselines.
For the baseline in the centre of the array, the cross takes on an `X' shape, since there are roughly the same number of antennas to the east of both antennas in the baseline as there are antennas to the west.
In terms of~\autoref{eq:coupled_vis_rewrite}, the positive delay terms come with both eastward-pointing baselines and westward-pointing baselines in roughly equal amounts, and the same is true for the negative delay terms.
The baseline at the edge of the array, however, presents a cross feature that takes on more of a `slash' shape, since the majority of the other antennas in the array are to the east of the antennas in the baseline.
Again referring to~\autoref{eq:coupled_vis_rewrite}, the positive delay terms mainly come with eastward-pointing baselines (corresponding to negative fringe-rates according to~\autoref{eq:f_peak}), while the negative delay terms mainly come with westward-pointing baselines.
Additionally, the relative strengths of the bar and cross features differ substantially between the two fields: the cross is prominent when a bright point source high in the sky dominates the field, while the bar is the main coupling feature when bright emission on the horizon dominates the field.

A slightly more nuanced point is that the LST 2.4--4.9 hr field also shows a prominent bar feature.
This feature is not affiliated with where the emission is coming from on the sky, but rather from the array geometry in the sense that most of the structure in the bar is due to coupling from North-South baselines.
Since North-South baselines are nearly parallel to the Earth's angular velocity vector,~\autoref{eq:fringe_rate} predicts that the bulk of the power in their visibilities is concentrated near zero fringe-rate, so these baselines will contribute to the bar feature almost regardless of where emission is sourced on the sky.
This is supported by inspecting the relative brightness of the bar feature between the two baselines shown in~\autoref{fig:compare_data_sim_fornax}, since fewer North-South baselines can be formed with the antennas at the edge of the array than with antennas at the centre of the array.
Though the dynamic range of the makes the comparison difficult, it appears that there is slightly more power in the bar for the baseline in the centre of the array than for the baseline at the edge of the array.

\begin{figure}
    \centering
    \includegraphics[width=\columnwidth]{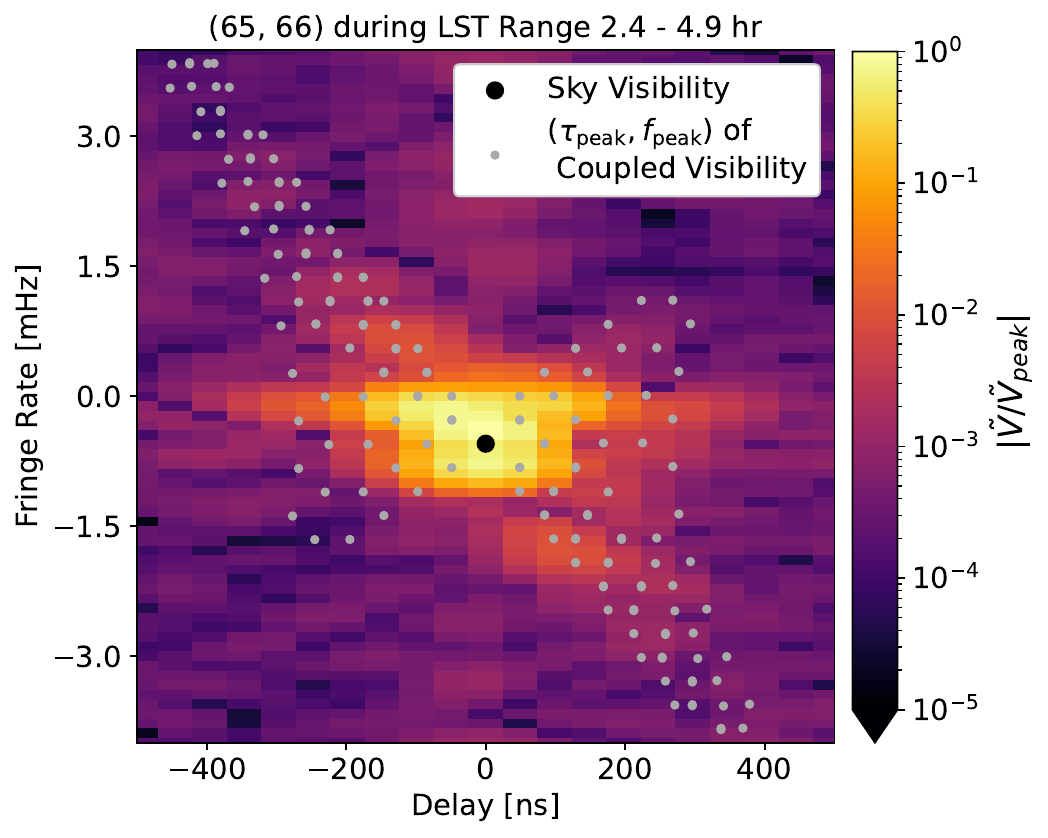}
    \caption{
        The regions of fringe-rate and delay space where the fringing portion of other visibilities should appear, according to our first-order coupling model.
        The data shown here are the same as in the bottom row of~\autoref{fig:compare_data_sim_fornax}.
        The grey dots are plotted according to the delays and fringe-rates computed with~\autoref{eq:tau_f_points}, with $\nu$ = 155 MHz, the central frequency of the band.
        Since the coupling amplitude is inversely proportional to distance, the contribution from antennas at higher delays (longer distances) falls below the noise level.
    }
    \label{fig:plot_with_dots}
\end{figure}

\subsection{The Amplitude of Coupled Visibilities}
\label{sec:AMPLITUDE}
We now return our attention to the discrepancies between the simulations and the data that were alluded to in Section~\ref{sec:COMPARISON}.
While the re-radiative first-order coupling model generates coupling features with a similar morphology to systematic features seen in the data, the simulated coupling features tend to be roughly an order-of-magnitude smaller than the features seen in the data.
The simulated amplitude of the coupled visibilities is set by the terms in $\mathbfss{X}$, as in \autoref{eq:coupling-coefficients}.
These terms encode information about the physics that produces mutual coupling and determine the amplitude at which it appears.
Therefore, any errors in modeling the physics of the re-radiation mechanism will manifest as incorrect predictions about the amplitude of the coupling features.




There is evidence to suggest that the primary mechanism by which mutual coupling is produced in HERA is different from the one examined in this paper.
Electromagnetic simulations of the HERA Phase I System by \citet{Fagnoni_2020} produce similar features in the visibility data, but find that most of the coupling structure is captured by interactions between the feed of one antenna and the dish of other antennas.
By contrast, re-radiation of the sky signal, as is the case in our first-order model, is a feed-to-feed interaction.
Since the geometry of the dish-to-feed coupling is similar to that of the feed-to-feed coupling, it is plausible that these two mechanisms would produce coupling features with similar morphologies in fringe-rate versus delay space.
The difference in the amplitude of the coupling between the first-order simulation and data may therefore be attributable to a discrepancy in the mechanism which produces the coupling.
Similar work is underway for the HERA Phase II system, and preliminary investigations corroborate this observation.
Alternatively, it is possible that the mechanism in \citet{Josaitis:2021a} is correct, but that errors in the beam model at the horizon are responsible for the mismatch in coupling amplitude between the data and the model. 

Regardless of whether the discrepancy in the amplitude of the coupling features is due to errors in the beam model or an alternative mechanism dominating the coupling, the fact remains that the morphology of the coupling features predicted by the re-radiative model is similar to the systematic excesses in power seen in the data.
We may therefore use~\autoref{eq:coupled_vis_rewrite} as a parametric description of mutual coupling and treat $\mathbfss{A}_{ij}$ as free parameters, which could then be fit to the data to remove mutual coupling features.
A successful strategy that addresses this form of coupling would be applicable to both the feed-to-feed interaction which we investigated here, as well as other mechanism for which astrophysical radiation redirected from neighboring antennas finds its way into the signal chain of other antennas. 
Such a strategy could be used to mitigate many different models of coupling, or even bypass the need to model the exact physics of coupling and propagation entirely.
Preliminary work employing this mitigation strategy is underway and will be explored further in a forthcoming publication.

\section{Mitigating Mutual Coupling with Fringe-Rate Filters}
\label{sec:MITIGATION}
In Section~\ref{sec:EVIDENCE}, we showed that data taken with the Phase II HERA instrument contain systematics that are in good qualitative agreement with predictions from our semi-analytic model of mutual coupling.
One of the key features of mutual coupling is the presence of excess power that manifests as a cross or bar feature in fringe-rate versus delay waterfalls (see e.g.,~\autoref{fig:compare_data_sim_fornax}).
In effect, the distribution of excess power due to mutual coupling manifests in fringe-rate in a predictable way, and this can be leveraged to mitigate its effects.

In this section, we describe how \emph{fringe-rate filtering} can be used to mitigate mutual coupling and use simulations to provide a cursory exploration of their efficacy and limitations.
Fringe-rate filtering, also referred to as delay-rate filtering, has a longstanding but relatively sparsely documented history as a systematics mitigation tool in radio astronomy, with the first application dating back at least as early as the 1970s (Bull, private communication).
~\citet{VSA_2003} used delay-rate filters similar to the `notch filter' described in Section~\ref{sec:NOTCH_FILTER} to mitigate slowly-varying `spurious signals' in tracking observations with the Very Small Array.
~\citet{Parsons_2009} extended the use of delay-rate filters to `select out' emission near an arbitrary phase centre for drift-scan observations, much like the `main lobe filter' discussed in Section~\ref{sec:MAIN-LOBE-FILTER}.
Additionally, as shown in~\citet{Charles:2023} and Charles et al. (in preparation), fringe-rate filters may be used prior to calibration to improve the quality of calibration solutions by improving array redundancy and reducing modeling errors associated with constructing absolute calibration models for widefield instruments.

Fundamentally, the application of a fringe-rate filter can be viewed as a time-averaging operation, and thus any fringe-rate filter will induce some level of signal loss and error covariance.
Correctly accounting for losses and covariances due to fringe-rate filtering turns out to be a nontrivial task, and detailed discussions are left to other works.
It is, however, possible to construct fringe-rate filters that retain some predetermined fraction of the cosmological signal, so the amount of signal loss incurred by fringe-rate filtering can be controlled.
An overview of the method for constructing such filters is provided in Section~\ref{sec:MAIN-LOBE-FILTER}.
A detailed treatment of filter-induced signal loss is provided in Pascua et al. (in preparation), and filter-induced error covariances are discussed in \citet{Wilensky_2023}.
A mathematically rigorous and generalised treatment of fringe-rate filters is provided in Martinot (in preparation).
These works are rounded out by a simulation-based investigation of how effects from fringe-rate filtering manifest in interferometric visibilities and power spectrum estimates in~\citet{Garsden_2024}.

\subsection{Notch Filter}
\label{sec:NOTCH_FILTER}
Recall from Section~\ref{sec:EVIDENCE} that the presence of bright emission on the horizon leads to excess `non-fringing' power.
This power tends to occupy just a few modes near zero fringe-rate, and thus may be excised from the data by applying a high-pass fringe-rate filter.
We refer to such a filter as a `notch' filter, since it notches out power at low fringe-rate, and it is effective at removing any sort of `common-mode' or non-fringing systematic.
Additionally, this type of filter can suppress horizon-delay foreground power that manifests from the pitchfork effect, and may prove useful for non-compact arrays that are less prone to mutual coupling effects.
It is important to note, however, that notch filters may come with undesirable side-effects, especially when applied to baselines that exhibit slow time variability in the visibilities (e.g., baselines with short East-West projected lengths or baselines on slowly rotating bodies such as the Moon).
Applying a notch filter to these baselines will not only remove non-fringing systematics, but will also incur substantial signal loss, as most of the emission from the sky occupies fringe-rate modes near zero for these baselines, as discussed in Section~\ref{sec:PREDICTIONS}.

\subsection{`Main-Lobe' Filter}
\label{sec:MAIN-LOBE-FILTER}
While the notch filter is a useful tool for mitigating the non-fringing features of mutual coupling, it typically leaves much of the coupling features unchanged.
The `main-lobe' filter is a more aggressive option for mitigating mutual coupling.
Motivated by~\citet{Parsons_2016}, one may design a fringe-rate filter which rejects features in the data that are inconsistent with emission from the main-lobe of the beam, effectively keeping only information that the interferometer is most sensitive to.
It turns out that a straightforward application of the formalism from~\citet{Parsons_2016} fails to generically produce fringe-rate filters that keep only the highest sensitivity regions of fringe-rate space; this approach only works in the limit that time variability in the data is dominated by sources drifting through the fringe pattern, and variability due to sources drifting through the primary beam is negligible.
Consequently, the fringe-rate filters applied in this analysis cannot exactly be interpreted as `beam-sculpting' operations; however, this interpretation is still useful as a conceptual device and motivates the `main-lobe filter' terminology.
For arrays like HERA, variability associated with sources drifting through the primary beam is non-negligible, and so we revisit the issue of properly characterizing time variability of visibilities in the fringe-rate domain.
A more complete treatment is given in Pascua et al. (in preparation); here we review the pertinent details.

Formally, the question we are trying to answer when designing `main-lobe' fringe-rate filters is the following:
For a given baseline, what is the correct characterization of the interferometer's intrinsic time variability in the fringe-rate domain?
In particular, we aim to design a filter that accepts the range of fringe-rate modes containing the majority of the cosmological 21-cm signal and rejects all other fringe-rate modes.
To that end, we simplify our analysis somewhat by switching to an unpolarised (or single-polarisation) treatment of the instrument response, since the cosmological 21-cm signal is expected to be mostly unpolarised~\citep{Babich_2005,Li_2021}.

In order to assess the instrument's \emph{intrinsic} variability, we consider its response to a sky-locked, flat spectrum Gaussian random field realization and compute the expected power in each fringe-rate mode.
We call this expected distribution of power in fringe-rate the \emph{fringe-rate profile}, and it can be computed via
\begin{equation}
    P(f_r) = \sum_m M_m \Biggl|\tilde{W}\biggl(f_r - \frac{m\omega_\oplus}{2\pi}\biggr)\Biggr|^2,
\label{eq:fringe-rate-profile}
\end{equation}
where $M_m$ is the \emph{$m$-mode power spectrum} and $W(t)$ is the tapering function applied before taking the fringe-rate transform, as previously noted in~\autoref{eq:fr-dly-transform}.
The $m$-mode power spectrum is computed as
\begin{equation}
    M_m = \sum_\ell \bigl|K_{\ell m}\bigr|^2,
\label{eq:m-mode-spectrum}
\end{equation}
where $K_{\ell m}$ is the \emph{beam transfer matrix}~\citep{Shaw_2014}, which is related to the primary beam and fringe pattern via
\begin{equation}
    A(\hat{\mathbfit{n}}) e^{-i2\pi\nu\mathbfit{b}\cdot\hat{\mathbfit{n}}/c} = \sum_{\ell,m} K_{\ell m} Y_\ell^m(\hat{\mathbfit{n}})^*,
\end{equation}
where $Y_\ell^m(\hat{\mathbfit{n}}) = P_\ell(\cos\theta) e^{im\phi}$ are spherical harmonics and $P_\ell(x)$ are Legendre polynomials.
The primary beam $A(\hat{\mathbfit{n}})$ is obtained from the product of Jones matrices $\mathbfss{JJ}^\dagger$, and the particular entry (or combination of entries) from this product depends on which visibility polarisation the filter is applied to: if the filters are to be applied to the co-polarisation (e.g., XX) or cross-polarisation (e.g., XY) terms, then $A(\hat{\mathbfit{n}})$ is simply one of the terms from $\mathbfss{JJ}^\dagger$; if the filters are to be applied to pseudo-Stokes polarisations, then $A(\hat{\mathbfit{n}})$ is obtained from a suitable linear combination of the terms in $\mathbfss{JJ}^\dagger$.
Note also that the convention for the beam transfer matrix used here differs from that of~\citet{Shaw_2014} in two ways: first, we use the peak-normalized primary beam rather than the integral normalized primary beam; second, we use the opposite sign convention for the fringe.

Since the $m$-mode power spectrum is computed entirely from quantities describing the instrument, it encodes the intrinsic time variability of the data.
Under periodic boundary conditions -- that is, a full sidereal day of observation -- the $m$-mode power spectrum exactly characterises the fringe-rate response of the instrument; however, real observations often only occupy a fraction of a sidereal day.
The incomplete sampling of a full day may be captured in the taper $W(t)$, and so the $\tilde{W}(f_r)$ term in~\autoref{eq:fringe-rate-profile} serves as a fringe-rate mode mixing matrix that accounts for non-periodic observations, as well as any tapering one might perform prior to applying the filter.

\begin{figure}
    \includegraphics[width=\columnwidth]{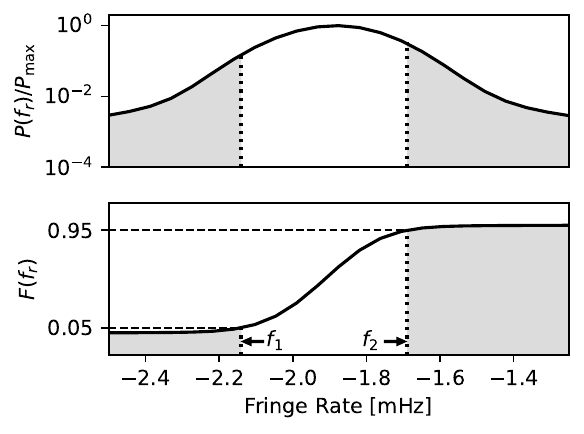}
    \caption{
        Schematic outlining how the main-lobe fringe-rate filter is constructed from a fringe-rate profile.
        The top panel shows an example fringe-rate profile, and the bottom panel shows the fringe-rate profile's cumulative distribution.
        The example fringe-rate profile shown here is computed for approximately five hours of observation, and so it showcases substantial wings due to the sinc sidelobes coming from the mode mixing matrix.
        The shaded grey regions indicate the regions of fringe-rate space rejected by the main-lobe filter defined by the filter bounds $(f_1, f_2)$.
    }
    \label{fig:fr_bound_schematic}
\end{figure}

\begin{figure}
    \includegraphics[width=\columnwidth]{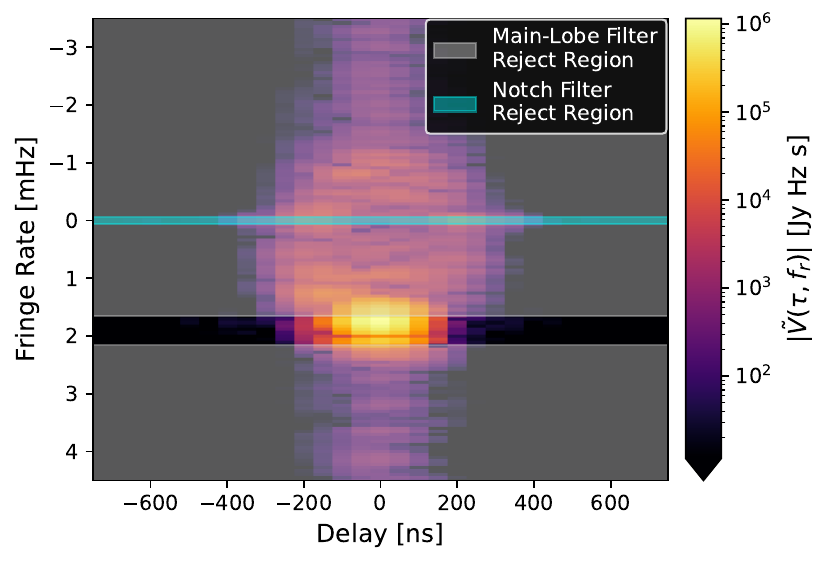}
    \caption{
        Qualitative schematic of how the main-lobe and notch filters operate on visibilities.
        The greyed out region indicates which fringe-rates are discarded by the main lobe filter, while the teal region indicates which fringe-rates are discarded by the notch filter.
        For this particular choice of baseline, the notch filter is degenerate with the main-lobe filter, but this degeneracy is broken on baselines with short East-West projected lengths, since the bulk of their power lies near zero fringe-rate.
    }
    \label{fig:filter_schematic}
\end{figure}

Given a fringe-rate profile, we may design a simple main-lobe filter by choosing some threshold of power to retain post-filtering, then integrating the fringe-rate profile to find suitable \emph{filter bounds} such that the desired level of power retention is met after applying a top-hat (i.e., bandpass) filter with these bounds.
The simplest approach to this is to first compute the normalized cumulative distribution of the fringe-rate profile,
\begin{equation}
    F(f_r) = \frac{\int_{-\infty}^{f_r} P(f_r') {\rm d}f_r'}{\int_\mathbb{R} P(f_r') {\rm d}f_r'},
\label{eq:fr-profile-cdf}
\end{equation}
then define percentile cuts $p_1, p_2$, where $p_2 - p_1$ is the power retained post-filtering.
Using this, the `main-lobe' fringe-rate filter bounds $f_1, f_2$ can be readily computed via $F(f_1) = p_1$ and $F(f_2) = p_2$, and we can finally define the main-lobe filter as
\begin{equation}
    \mathcal{F}(f_r) = \begin{cases}
        1, \quad f_1 \leq f_r \leq f_2 \\
        0, \quad {\rm otherwise}
    \end{cases}.
\label{eq:main-lobe-filter}
\end{equation}
A schematic of the procedure used to compute the main-lobe filter bounds is shown in~\autoref{fig:fr_bound_schematic}.
A visualization of the main-lobe and notch filters in fringe-rate versus delay space is shown in~\autoref{fig:filter_schematic}.

\subsection{Application to Simulations}
In order to test the efficacy of the main-lobe filter, we simulate realistic visibilities and apply first-order coupling as described in Section~\ref{sec:SIMULATION}, then filter the coupled visibilities according to the filtering scheme described above.
In this section, we briefly describe the analysis performed to compute the power spectrum results presented in Section~\ref{sec:POWER-SPECTRUM}.

\subsubsection{Filter Implementation}
\label{sec:FILTER_IMPLEMENTATION}
Fringe-rate filtering was performed baseline-by-baseline on the five hours of simulation data used for the comparisons performed in Section~\ref{sec:EVIDENCE}.
The filter bounds for each baseline were computed based on the overview provided in the previous section, but the application of this technique to data requires some care to avoid inducing extra spectral structure on the filtered data.
A straightforward application is to compute a different fringe-rate profile for each frequency, derive filter bounds from that profile, then proceed with the filtering on a frequency-by-frequency basis; however, although the fringe-rate profiles evolve smoothly in frequency, applying a filter that varies over the subband used for power spectrum estimation could potentially add spectral structure to the filtered data and reduce the filter's utility as a coupling mitigation tool.
In order to avoid this potential complication, we first compute fringe-rate profiles for each frequency in the subband used for analysis, then average these together with the same tapering function used in power spectrum estimation.
This averaged fringe-rate profile is then used to derive the filter bounds.
We use 5\% and 95\% percentile cuts for obtaining the filter bounds.

Rather than apply the filters directly in the fringe-rate domain, we opt to instead fit the data to a model built from Discrete Prolate Spheroidal Sequences (DPSS)~\citep{Slepian_1978}, which are band-limited in both time and fringe-rate simultaneously.
In addition to their spectral concentration properties, filtering by fitting to the DPSS basis allows us to prevent numerical artefacts associated with Fourier transform sidelobes from contaminating the filtered data.
Each \emph{DPSS mode} $\phi_n(t)$ is a function that can be characterised by its spectral concentration, or the fraction of power contained within the filter bounds, and is uniquely determined by a half-bandwidth parameter and the number of samples in the time series.
The filter is applied by removing DPSS modes whose spectral concentration is below some threshold value and fitting the data to the remaining DPSS modes.
By keeping only modes with a sufficient fraction of their power within the filter bounds and fitting the data to these modes, we are effectively filtering the data to only keep the portion of the signal occupying fringe-rates within the filter bounds.
As the number of retained modes increases, the sharpness of the filter also increases, but at the expense of increased spillover beyond the filter bounds.

\begin{figure*}
    \includegraphics[width=\textwidth]{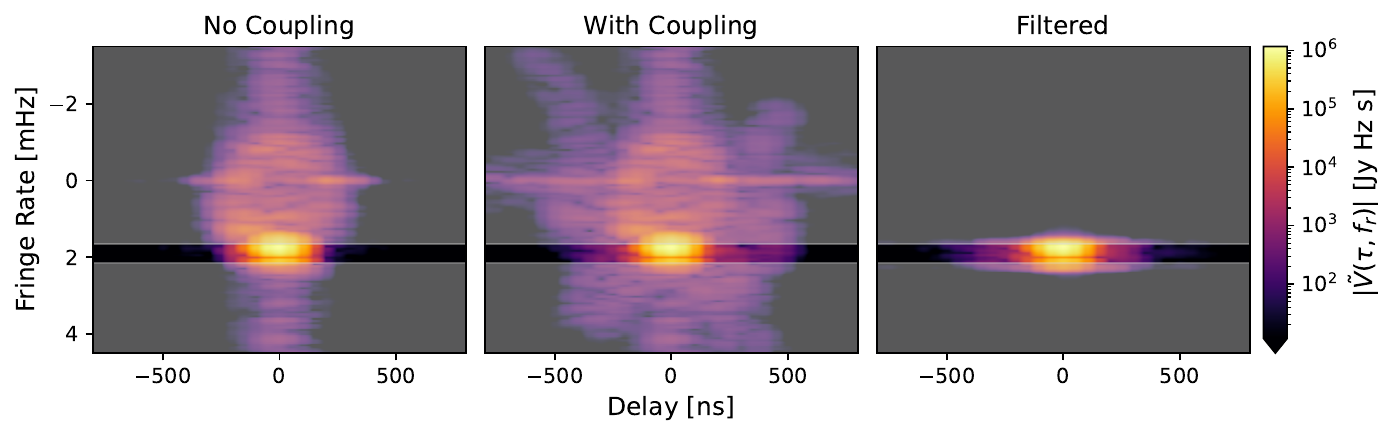}
    \caption{
        Comparison of the uncoupled, coupled, and filtered data in fringe-rate versus delay space for a 58-m East-West baseline.
        The greyed out regions indicate the fringe-rates discarded by the main lobe filter that was used to produce the data in the rightmost panel.
        While there is a small amount of signal that remains just beyond the filter boundaries, the majority of the signal, and thus the majority of the mutual coupling, has been completely removed by the filter.
    }
    \label{fig:filtered-data-waterfalls}
\end{figure*}

Formally, we are modeling the data via
\begin{equation}
    V_{ij}^{\rm filt}(t) = \sum_n a_n \phi_n(t)
\end{equation}
and performing a least-squares fit to determine $a_n$.
For the test case investigated here, we use a permissive filter by setting the spectral concentration threshold to $10^{-9}$.
An example of how the filter affects the visibilities is visualized in fringe-rate versus delay space in~\autoref{fig:filtered-data-waterfalls}.
A pedagogical review of filtering with DPSS modes is provided in HERA Memo 129~\citep{DPSS_Memo}\footnote{\url{https://reionization.org/manual_uploads/HERA129_DPSS_fringe_rate_filter_implementation_v1.pdf}}.

\subsubsection{Power Spectrum Estimation}
\label{sec:PSPEC-ESTIMATION}
Visibilities are propagated to power spectrum estimates with the delay spectrum formalism~\citep{Parsons_2012}, which is implemented in \textsc{hera\_pspec}\footnote{\url{https://github.com/HERA-Team/hera_pspec}} with the quadratic estimator formalism described in~\citet{HERA_Limits_2022}.
In the delay spectrum formalism, one first computes the delay transform of the visibilities via
\begin{equation}
    \tilde{V}_{ij}(\tau,t) = \int T(\nu) V_{ij}(\nu,t) e^{-i2\pi\nu\tau} {\rm d}\nu,
\end{equation}
where $\tau$ is delay and $T(\nu)$ is a tapering function applied to suppress sidelobes from the sharp cutoff of the foreground signal at the band edges.
Delay spectra $\tilde{V}_{ij}(\tau,t)$ are converted to probes of cosmological Fourier modes $\tilde{V}(\vec{k}_\perp, k_\parallel, t)$ through the correspondences~\citep{Hogg_2000,Liu:2014a}
\begin{align}
    \mathbfit{k}_\perp &= \frac{2\pi}{Y} \frac{\mathbfit{b}}{\lambda}, \\
    k_\parallel &= \frac{2\pi\tau}{X},
\end{align}
where $k_\parallel$ is a cosmological Fourier mode along the line-of-sight, $\mathbfit{k}_\perp$ is a cosmological Fourier mode in the plane transverse to the line-of-sight, and $X$ and $Y$ are cosmological scalars that convert frequencies and sky angles to cosmological distances, respectively.
The cosmological scalars $X$ and $Y$ are defined via
\begin{align}
    X &= \frac{c(1+z)^2}{\nu_{21} H(z)}, \\
    Y &= D(z),
\end{align}
where $D(z)$ is the transverse comoving distance, $H(z)$ is the Hubble parameter, $\nu_{21} \approx 1420\ {\rm MHz}$ is the rest frequency of the 21-cm line, and the redshift $z$ is taken at the centre of the frequency band used in the delay transform.
Power spectra are then estimated via
\begin{equation}
    \hat{P}(\mathbfit{k}_\perp, k_\parallel) = \frac{X^2 Y}{B\Omega_{\rm pp}} \Bigl\langle \bigl|\tilde{V}(\mathbfit{k}_\perp, k_\parallel, t)\bigr|^2 \Bigr\rangle_t,
\end{equation}
where $B$ is the effective bandwidth over which the delay transform was performed, $\Omega_{\rm pp}$ is the integral of the square of the primary beam, and $\langle \cdot \rangle_t$ indicates a time average.
In the delay transform we weight by a 7-term Blackman-Harris tapering function~\citep{Blackman_1958}, and we additionally coherently average the visibilities from all baselines within a redundant group (i.e., over all identical baselines) prior to power spectrum estimation.
Cylindrically and spherically averaged power spectrum estimates are shown in~\autoref{fig:power-spectrum-plots}; when performing the averages, we exclude baselines with an East-West projected length less than 15 meters.
We additionally exclude modes within the `foreground wedge' when performing the spherical average by ignoring delay modes less than the geometric horizon (the maximum geometric delay for emission from the sky, ~\citealt{Parsons_2012}) plus a 325 ns buffer.

\section{Effects of Mutual Coupling on Power Spectrum Estimation}
\label{sec:POWER-SPECTRUM}

\begin{figure*}
    \includegraphics{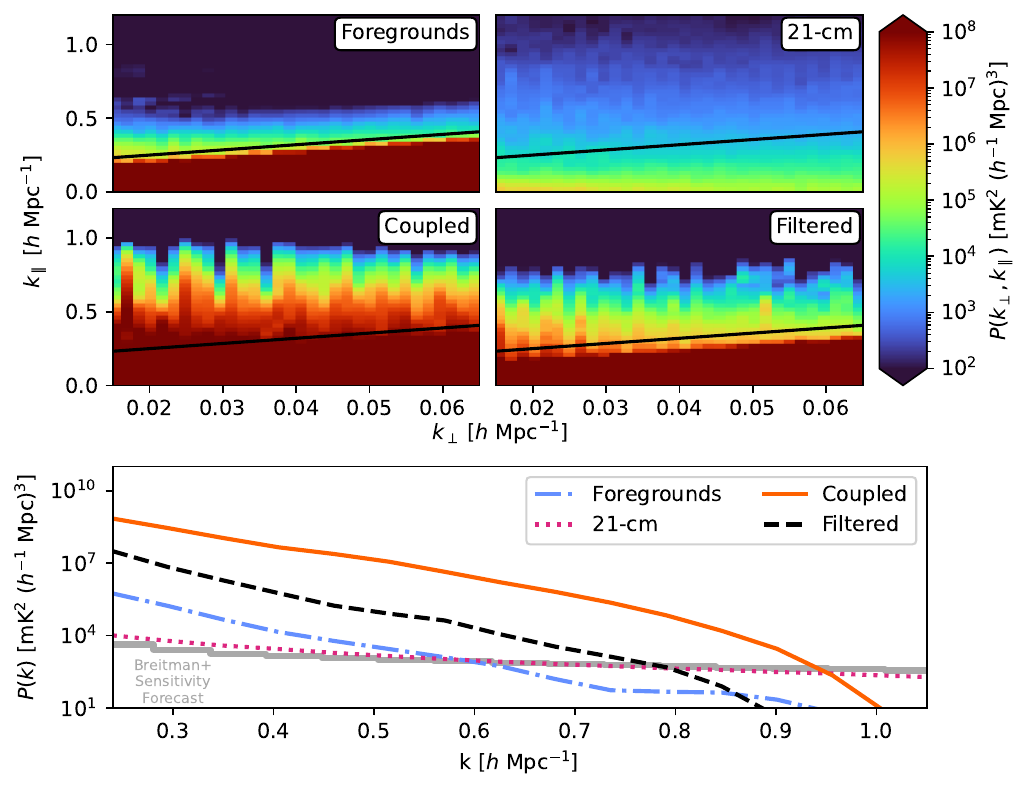}
    \caption{
        Comparison of simulated power spectrum estimates averaged over the LST 0--2.4 hour field, only including baselines with an East-West projection longer than 15 meters.
        The group of four cylindrical power spectra in the upper portion of the figure show: foregrounds in the upper left; 21-cm signal in the upper right; foregrounds with first-order coupling in the bottom left; a coupled signal that has been fringe-rate filtered in the bottom right.
        The black lines indicate the `wedge cut' that was used when forming spherically averaged power spectra, where only modes above the black line were included in the averaging.
        The bottom panel shows the spherically averaged spectra obtained from the upper four panels, alongside the forecast sensitivity for the full $\sim$140 nights of data from the sixth HERA observing season.
        This figure shows that first-order mutual coupling effects contaminate the EoR window; in other words, modes that would otherwise be dominated by the cosmological 21-cm signal are obscured by mutual coupling.
        Fringe-rate filtering helps mitigate the effects of mutual coupling and recovers a substantial portion of the EoR window, but it does not solve the problem entirely.
        This is most clearly seen in the spherically averaged spectra, where the transition to the 21-cm dominated region occurs at $k \sim 0.95\ h\ {\rm Mpc}^{-1}$ for the coupled case, $k \sim 0.8\ h\ {\rm Mpc}^{-1}$ for the filtered case, and $k \sim 0.6\ h\ {\rm Mpc}^{-1}$ for the uncoupled case.
    }
    \label{fig:power-spectrum-plots}
\end{figure*}

Mutual coupling poses serious issues for detecting the 21-cm power spectrum, as evidenced by the simulated power spectrum estimates shown in~\autoref{fig:power-spectrum-plots}.
Comparing the upper two panels of~\autoref{fig:power-spectrum-plots}, we see that in the absence of mutual coupling, there are many $(\mathbfit{k}_\perp, k_\parallel)$ modes that are dominated by the cosmological 21-cm signal; this is the EoR window targeted by foreground-avoidance approaches to detecting the 21-cm signal.
The middle two panels in~\autoref{fig:power-spectrum-plots} paint a grim picture: a large swathe of the EoR window is corrupted by mutual coupling, and fringe-rate filtering seems to only mildly alleviate the issue.
In addition to this,~\autoref{fig:power-spectrum-plots} shows that the filtering scheme proposed in Section~\ref{sec:FILTER_IMPLEMENTATION} is less effective for short baselines than it is for long baselines.
This, however, is not an unexpected result, since most of the power is concentrated closer to zero fringe-rate for shorter baselines.
Consequently, more of the bar feature associated with mutual coupling is leftover after fringe-rate filtering on short baselines and hence these baselines retain more mutual coupling than long baselines do after filtering.

Clearly, mutual coupling is a serious obstacle that must be overcome in order to detect the cosmological 21-cm signal.
Despite the substantial suppression obtained by fringe-rate filtering the data, which is certainly helpful in the context of setting upper limits on the 21-cm power spectrum,~\autoref{fig:power-spectrum-plots} suggests that fringe-rate filters simply do not perform well enough to recover a detection close to the foreground wedge.
The bottom panel of~\autoref{fig:power-spectrum-plots} accentuates the urgency and importance of this point.
In thick grey, we reproduce the forecasted error bars on a HERA power spectrum measurement from~\citet{Breitman:2023}.
As shown in~\citet{Breitman:2023}, from the standpoint of sensitivity alone, the sixth season of HERA data is capable of detecting the cosmological 21-cm signal with order unity signal-to-noise in each $k$-bin, provided systematics in the data can be brought under control.
It is therefore a non-negotiable requirement that we develop better mutual coupling mitigation tools if we intend to unlock the full design potential of HERA.

Rather than lament at the challenge presented by the results in~\autoref{fig:power-spectrum-plots}, we ought to use the tools we have available to better understand the features in~\autoref{fig:power-spectrum-plots}.
Recall that the power spectrum estimation method employed is effectively just a `Fourier transform and square' method, and a crucial assumption is that visibilities are treated as probes of individual transverse cosmological Fourier modes.
The first-order coupling model is enough to show that mutual coupling strongly breaks this assumption, since in the first-order model each coupled visibility is effectively a linear combination of the uncoupled visibilities.
Taking the delay transform of the unpolarized version of~\autoref{eq:first-order-visibility}, we obtain
\begin{equation}
    \label{eq:delay-transformed-coupling}
    \tilde{V}_{ij}^{(1)}(\tau) = \tilde{V}_{ij}^{(0)}(\tau) + \sum_k \Bigl[\tilde{V}_{ik}^{(0)} * \tilde{X}_{jk}^*\Bigr](\tau) + \Bigl[\tilde{X}_{ik} * \tilde{V}_{kj}^{(0)}\Bigr](\tau),
\end{equation}
where $\tilde{A} * \tilde{B}$ indicates a convolution in delay.
Three important conclusions can be made from~\autoref{eq:delay-transformed-coupling}: First, in the presence of mutual coupling, no baseline measures a single $k_\perp$ mode, but rather each $k_\perp$ bin contains contributions from all baselines in the array; Second, as a corollary to the first point, since the coupled delay spectra contain contributions from all of the uncoupled delay spectra, the foreground wedge is effectively smeared out in $k_\perp$; Third, since the coupling coefficients have nontrivial frequency structure, the convolution of the uncoupled delay spectra with the `coupling kernel' $\tilde{X}$ introduces an additional smearing in $k_\parallel$.
To put it another way, mutual coupling causes the chromaticity of the entire array to be imprinted on each baseline in the array, and this effect is compounded by additional delay smearing induced by the frequency-dependent evolution of the coupling coefficients.
Mutual coupling does not act to simply smear out foreground power strictly horizontally or vertically in the $(k_\perp, k_\parallel)$ plane, but rather simultaneously smears out foreground power in both $k_\perp$ and $k_\parallel$.

As a closing note, we provide some predictions for how effects ignored in the first-order model could manifest in power spectrum estimates.
A simple conceptual extension of the first-order model is to allow for higher-order coupling terms by considering the correlation of re-radiated signals with each other (i.e., keeping the terms that were discarded in the first-order treatment) and extending the model to include multiple instances of re-radiation.
Both of these extensions of the formalism predict additional coupling power with complex delay structure, some of which would survive the main-lobe filter.
In effect, extensions of the first-order model would predict that the coupling effects extend to even higher $k_\parallel$ modes than seen in~\autoref{fig:power-spectrum-plots}, and less suppression in the regions that clearly benefited from the main-lobe filter.
Consequently, as experiments creep closer toward detecting the cosmological 21-cm signal, analysts will need to employ increased caution when interpreting power spectrum estimates, as untreated higher-order coupling effects could potentially masquerade as the cosmological 21-cm signal.

\section{Conclusion}
\label{sec:CONCLUSION}
In this work, we analyze HERA Phase II data for features indicative of mutual coupling between array elements. 
We compare HERA Phase II data to simulations of the same array configuration and observational parameters (summarised in~\autoref{table:obs_params}), using unpolarised skies with diffuse and point-source emission.
We find qualitative agreement between the fringing and non-fringing coupling features, with the brightest coupling features appearing at the order of a few parts in $10^3$ relative to the peak visibility amplitude.
We find that the coupling features predicted by the first-order re-radiative coupling model systematically under-predict the amplitude of the coupling features, which may either come from errors in the beam model at the horizon or may be indicative of the re-radiative mechanism being sub-dominant to other coupling mechanisms.
In Section~\ref{sec:MITIGATION}, we described a fringe-rate filtering strategy designed to mitigate this type of coupling, and found in Section~\ref{sec:POWER-SPECTRUM} that the proposed fringe-rate filtering scheme helps ameliorate the issue, but does not completely remove coupling signatures.
Given the good qualitative agreement between the simulated coupling features and systematic effects seen in the data, we posit that it may be possible to further mitigate mutual coupling by reparametrising the coupling coefficients as $\mathbfss{X}_{ij} = \mathbfss{A}_{ij} \exp(i2\pi\nu\tau_{ij})$ and fitting for $\mathbfss{A}_{ij}$.

We found that the main-lobe fringe-rate filtering technique discussed in Section~\ref{sec:MAIN-LOBE-FILTER} helps mitigate the effects of mutual coupling.
We assessed the efficacy of this filtering scheme by simulating first-order coupling, applying a main-lobe fringe-rate filter to the simulated data, and comparing power spectrum estimates between the filtered and unfiltered simulation data.
We found that the simulated coupling features completely dominated the EoR window, and that applying the main-lobe fringe-rate filter recovered a modest fraction of the EoR window lost to coupling.
We thus conclude that the main-lobe fringe-rate filter certainly helps alleviate the issues posed by mutual coupling, but improved mitigation techniques are required to realise the full potential of HERA.
Compounding this with our findings that the simulated coupling is roughly an order-of-magnitude fainter than what is seen in the data accentuates the issue: mutual coupling poses a serious threat to a densely packed array's ability to detect the cosmological 21-cm signal, and better removal techniques need to be developed in order to deal with this problem.

In closing, although mutual coupling is a significant issue that needs to be addressed, the situation is not beyond salvaging.
New mitigation and subtraction techniques are currently being developed, and the filtering techniques proposed here only scratch the surface of what is possible.
More general parametrisations of the coupling may provide a promising route to gaining increased suppression of coupling features, and it is possible that signal processing techniques used in other applications, such as adaptive filters, could serve as useful mitigation tools.
Since the first-order coupling model, which is computationally cheap relative to full electromagnetic simulations, predicts coupling features that are in very good qualitative agreement with what is seen in the data, analysts may use first-order coupling simulations to test their mitigation techniques without needing to tackle up-front all of the additional challenges that come with real data.
In addition to finding software-based solutions, next-generation interferometers designed to detect the 21-cm signal from CD and the EoR can address this with hardware-based solutions by carefully designing their arrays so that coupling features are naturally suppressed, and some upcoming experiments have already begun taking this approach~\citep{hirax-design,vincent-thesis}.
Alternatively, another approach is to better understand the hardware as-is by mapping out the coupled beam through a data-driven approach, as is done by CHIME~\citep{CHIME:2022}.
The current situation is challenging, but there is hope on the horizon.

\section*{Acknowledgements}
We thank Vincent MacKay, Bang Nhan, and Jonathan Sievers for discussions which contributed to this work.
This material is based upon work supported by the National Science Foundation under Grant Nos. 1636646, 1836019, AST-2144995 and institutional support from the HERA collaboration partners.
This research is funded in part by the Gordon and Betty Moore Foundation through Grant GBMF5212 to the Massachusetts Institute of Technology.
This work was funded in part by the Canada 150 Research Chairs Program.
HERA is hosted by the South African Radio Astronomy Observatory, which is a facility of the National Research Foundation, an agency of the Department of Science and Innovation.

AL and RP acknowledge support from the Trottier Space Institute, the Canadian Institute for Advanced Research (CIFAR) Azrieli Global Scholars program, a Natural Sciences and Engineering Research Council of Canada (NSERC) Discovery Grant, a NSERC/Fonds de recherche du Québec -Nature et Technologies NOVA grant, the Sloan Research Fellowship, and the William Dawson Scholarship at McGill.
EdLA is supported by an STFC Ernest Rutherford Fellowship.
NK acknowledges support from NASA through the NASA Hubble Fellowship grant \#HST-HF2-51533.001-A awarded by the Space Telescope Science Institute, which is operated by the Association of Universities for Research in Astronomy, Incorporated, under NASA contract NAS5-26555.
This result is part of a project that has received funding from the European Research Council (ERC) under the European Union's Horizon 2020 research and innovation programme (Grant agreement No. 948764; PB, JB, MJW).
SGM has received funding from the European Union's Horizon 2020 research and innovation programme under the Marie Skłodowska-Curie grant agreement No 101067043.
YZM is supported by the National Research Foundation of South Africa with Grant No. 150580, No. 159044, No. CHN22111069370 and No. ERC23040389081.

\section*{Software and Data Availability}
The data products used in this analysis may be requested via email from the authors.
Information on the calibration and LST-binning routines are found in HERA Memo 124~\citep{IDR21_Memo}.
Visibility data was managed using \textsc{pyuvdata}\footnote{\url{https://github.com/RadioAstronomySoftwareGroup/pyuvdata}}.
Inpainting and fringe-rate filtering was performed using tools available in \textsc{hera\_cal}\footnote{\url{https://github.com/HERA-Team/hera_cal}}.
Power spectrum estimation was performed using tools available in \textsc{hera\_pspec}.
Visibility simulation was performed with \textsc{matvis}\footnote{\url{https://github.com/HERA-Team/matvis}}~\citep{Kittiwisit_2023} using wrappers provided by \textsc{hera\_sim}.
A demonstration of how to apply the first-order coupling formalism to interferometric visibility data is provided in a tutorial notebook in the \textsc{hera\_sim} repository.

\bibliographystyle{mnras}
\bibliography{main} 


\appendix

\section{Working Through the First Order Coupling Model}
\label{sec:appendix}

We examine the case of a three-element array in order to gain intuition for the first-order mutual coupling equations.
\autoref{fig:3_ant_array_layout} shows the array layout.
In Section~\ref{sec:worked_example}, we explicitly compute \autoref{eq:first-order-visibility} for the example array, starting with the electric fields incident upon each antenna.
The three-element array is also instructive for understanding the behavior of coupling in delay and fringe rate space, as the small number of elements allows us to isolate the power from each coupled antenna.
Section~\ref{sec:FRATE-DELAY-TRANSFORM} quickly reviews the delay and fringe rate transforms, and simulates first-order coupling for the three antenna array using a toy visibility model.

\subsection{A Worked Example of the Coupling Equation}
\label{sec:worked_example}

\begin{figure}
    \centering
    \includegraphics[width=\columnwidth]{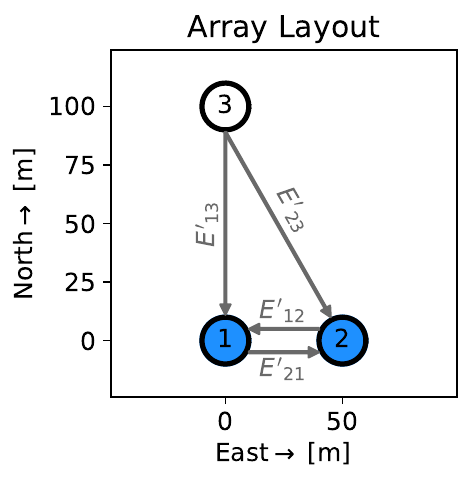}
    \caption{
        The layout for the 3-element array we will consider, as well as the re-radiated electric fields from each antenna.
        Antenna 1 is located at (0, 0) meters, antenna 2 at (50, 0) meters, and antenna 3 at (0, 100) meters.
    }
    \label{fig:3_ant_array_layout}
\end{figure}

In this section, we will examine the coupling in the visibility for the baseline formed by antennas 1 and 2 in~\autoref{fig:3_ant_array_layout}.
The visibility is formed by correlating the voltages produced by each antenna's signal chain, $\mathbfit{v}_1^{\rm tot}$ and $\mathbfit{v}_2^{\rm tot}$, which gives 
\begin{equation}
\label{eq:V_12}
    \mathbfss{V}_{12}^{(1)} = \Bigl\langle \mathbfit{v}_1^{\rm tot} \bigl(\mathbfit{v}_2^{\rm tot}\bigr)^\dag \Bigr\rangle.
\end{equation}
We will insert~\autoref{eq:measured-voltage} into the above equation and explicitly evaluate all of the resulting terms for the three element array shown in~\autoref{fig:3_ant_array_layout}.

As described in Section~\ref{sec:COUPLED-VISIBILITIES}, mutual coupling occurs because a part of the incident electric field on each antenna is re-radiated to other antennas.
$\mathbfit{v}_1^{\rm tot}$ and $\mathbfit{v}_2^{\rm tot}$ therefore contain terms which correspond to both the sky signal and the re-radiated signal from mutually coupled antennas.
\autoref{eq:scattered-field-two-antennas} describes the form of the electric field re-radiated from antenna $j$ into antenna $i$,  $\mathbfit{E}'_{ij}(\hat{\mathbfit{n}})$.
Making the substitution in \autoref{eq:coeff-sub}, and taking $i \neq j$,  $\mathbfit{E}'_{ij}$ takes the form
\begin{equation}
\label{eq:E_scatter}
    \mathbfit{E}'_{ij}(\hat{\mathbfit{n}}) = \frac{i \Gamma}{\Omega_{\rm p}} \frac{e^{2 \pi i \nu \tau_{ij}}}{u_{ij}} \delta^{\rm D}(\hat{\mathbfit{n}} - \hat{\mathbfit{b}}_{ij}) \mathbfss{J}(\hat{\mathbfit{b}}_{ji})^\dag \int_{4\pi} \mathbfss{J}(\hat{\mathbfit{n}}') \mathbfit{E}_j(\hat{\mathbfit{n}}') {\rm d}\Omega',
\end{equation}
where $u_{ij} = b_{ij} / \lambda$ is the baseline length in units of wavelengths.
Antenna 1 will receive re-radiated sky signal from antenna 2, $\mathbfit{E}'_{12}$, as well as antenna 3, $\mathbfit{E}'_{13}$ , in addition to its own incident sky signal $\mathbfit{E}_1$.
Likewise, antenna 2 will observe contributions from both antenna 1, $\mathbfit{E}'_{21}$, and antenna 3, $\mathbfit{E}'_{23}$, in addition to its own sky signal $\mathbfit{E}_2$.
The re-radiated fields and their propagation paths are labeled in \autoref{fig:3_ant_array_layout}.
The complete form of the E-field measured by both antennas are thus

\begin{equation}
\begin{aligned}
    \mathbfit{E}_1^{\rm tot}(\hat{\mathbfit{n}}) = \mathbfit{E}_1(\hat{\mathbfit{n}}) + \mathbfit{E}'_{12}(\hat{\mathbfit{n}}) 
    + \mathbfit{E}'_{13}(\hat{\mathbfit{n}}), \\
    \mathbfit{E}_2^{\rm tot}(\hat{\mathbfit{n}}) = \mathbfit{E}_2(\hat{\mathbfit{n}}) + \mathbfit{E}'_{21}(\hat{\mathbfit{n}}) 
    + \mathbfit{E}'_{23}(\hat{\mathbfit{n}}).
\end{aligned}
\end{equation}

$\mathbfit{E}^{\rm tot}_1$ and $\mathbfit{E}^{\rm tot}_2$ will be read out of their respective signal chains as voltages according to \autoref{eq:measured-voltage}, with $\mathbfit{E}_i$ replaced with $\mathbfit{E}^{\rm tot}_i$.
For antennas 1 and 2, the voltages $\mathbfit{v}_1$ and $\mathbfit{v}_2$ are

\begin{equation}
\begin{aligned}
    \mathbfit{v}_1^{\rm tot} = h_0 \int_{4\pi} \mathbfss{J}(\hat{\mathbfit{n}}) \left(\mathbfit{E}_1(\hat{\mathbfit{n}}) + \mathbfit{E}'_{12}(\hat{\mathbfit{n}}) + \mathbfit{E}'_{13}(\hat{\mathbfit{n}}) \right) {\rm d} \Omega, \\
    \mathbfit{v}_2^{\rm tot} = h_0 \int_{4\pi} \mathbfss{J}(\hat{\mathbfit{n}}) \left(\mathbfit{E}_2(\hat{\mathbfit{n}}) + \mathbfit{E}'_{21}(\hat{\mathbfit{n}}) + \mathbfit{E}'_{23}(\hat{\mathbfit{n}}) \right) {\rm d} \Omega.
\end{aligned}
\end{equation}

We now perform the integral over solid angle ${\rm d} \Omega$.
We can substitute $\mathbfit{v}_i$ after integrating the first term in both equations.
This is the voltage which antenna $i$ would have measured in the absence of any mutual coupling.
The presence of the delta function in~\autoref{eq:E_scatter} just picks out the $\hat{\mathbfit{n}} = \hat{\mathbfit{b}}_{ij}$ term when computing the integral for the re-radiated electric field terms.
The equations for the voltages then become

\begin{align}
    \nonumber
    \mathbfit{v}_1^{\rm tot} &= \mathbfit{v}_1 + \frac{i \Gamma}{\Omega_{\rm p}} \frac{e^{i2 \pi \nu \tau_{21}}}{u_{21}} h_0 \mathbfss{J}(\hat{\mathbfit{b}}_{12}) \mathbfss{J}(\hat{\mathbfit{b}}_{21})^\dag \int \mathbfss{J}(\hat{\mathbfit{n}}') \mathbfit{E}_2(\hat{\mathbfit{n}}') {\rm d} \Omega' \\
    \label{eq:v_1_tot}
    &\hspace{0.66cm} + \frac{i \Gamma}{\Omega_{\rm p}} \frac{e^{i2 \pi \nu \tau_{31}}}{u_{31}} h_0 \mathbfss{J}(\hat{\mathbfit{b}}_{13}) \mathbfss{J}(\hat{\mathbfit{b}}_{31})^\dag \int \mathbfss{J}(\hat{\mathbfit{n}}') \mathbfit{E}_3(\hat{\mathbfit{n}}') {\rm d} \Omega', \\
    \nonumber
    \mathbfit{v}_2^{\rm tot} &= \mathbfit{v}_2 + \frac{i \Gamma}{\Omega_{\rm p}} \frac{e^{i2 \pi \nu \tau_{12}}}{u_{12}} h_0 \mathbfss{J}(\hat{\mathbfit{b}}_{21}) \mathbfss{J}(\hat{\mathbfit{b}}_{12} )^\dag \int \mathbfss{J}(\hat{\mathbfit{n}}) \mathbfit{E}_1(\hat{\mathbfit{n}}) {\rm d} \Omega' \\
    \label{eq:v_2_tot}
    &\hspace{0.66cm} + \frac{i \Gamma}{\Omega_{\rm p}} \frac{e^{i2 \pi \nu \tau_{32}}}{u_{32}} h_0 \mathbfss{J}(\hat{\mathbfit{b}}_{23}) \mathbfss{J}(\hat{\mathbfit{b}}_{32})^\dag \int \mathbfss{J}(\hat{\mathbfit{n}}) \mathbfit{E}_3(\hat{\mathbfit{n}}) {\rm d} \Omega'.
\end{align}
The above equations contain terms of $\mathbfss{X}$ as in \autoref{eq:coupling-coefficients}. 
We again make use of \autoref{eq:measured-voltage} when taking the integral over ${\rm d} \Omega'$.
The resulting voltages measured by antenna 1 and antenna 2 are 

\begin{equation}
\begin{aligned}
    \mathbfit{v}_1^{\rm tot} = \mathbfit{v}_1 + \mathbfss{X}_{12} \mathbfit{v}_2 + \mathbfss{X}_{13} \mathbfit{v}_3, \\
    \mathbfit{v}_2^{\rm tot} = \mathbfit{v}_2 + \mathbfss{X}_{21} \mathbfit{v}_1 + \mathbfss{X}_{23} \mathbfit{v}_3.
\end{aligned}
\end{equation}

This form of the voltages makes it easier to identify coupled visibilities in the full equation for $\mathbfss{V}_{12}^{(1)} = \langle \mathbfit{v}_1^{\rm tot} (\mathbfit{v}_2^{\rm tot})^\dag\rangle$.
We emphasize again here that the coupling is caused by copies of the \textit{electric field} being re-radiated.
Keeping only terms to first order, we correlate the voltages to obtain the following expression for $\mathbfss{V}_{12}^{(1)}$ 

\begin{align}
    \nonumber
    \mathbfss{V}_{12}^{(1)} &= \Bigl\langle \mathbfit{v}_1 \mathbfit{v}_2^\dag \Bigr\rangle + \Bigl\langle \bigl(\mathbfss{X}_{12} \mathbfit{v}_2\bigr) \mathbfit{v}_2^\dag\Bigr\rangle + \Bigl\langle \bigl(\mathbfss{X}_{13} \mathbfit{v}_3\bigr) \mathbfit{v}_2^\dag\Bigr\rangle \\ 
    \nonumber
    &\hspace{0.275cm} + \Bigl\langle \mathbfit{v}_1 \bigl(\mathbfss{X}_{21} \mathbfit{v}_1\bigr)^\dag \Bigr\rangle + \Bigl\langle \mathbfit{v}_1 \bigl(\mathbfss{X}_{23} \mathbfit{v}_3\bigr)^\dag \Bigr\rangle \\
    \nonumber
    &= \langle \mathbfit{v}_1 \mathbfit{v}_2^\dag \rangle + \mathbfss{X}_{12} \langle \mathbfit{v}_2 \mathbfit{v}_2^\dag\rangle + \mathbfss{X}_{13} \langle \mathbfit{v}_3 \mathbfit{v}_2^\dag\rangle + \langle \mathbfit{v}_1 \mathbfit{v}_1^\dag \rangle \mathbfss{X}_{21}^\dag  + \langle \mathbfit{v}_1 \mathbfit{v}_3^\dag \rangle \mathbfss{X}_{23}^\dag \\
\label{eq:3_ant_coupling_eq}
    &= \mathbfss{V}_{12}^{(0)} + \mathbfss{X}_{12} \mathbfss{V}_{22}^{(0)} + \mathbfss{X}_{13} \mathbfss{V}_{32}^{(0)}
    + \mathbfss{V}_{11}^{(0)}\mathbfss{X}_{21}^\dag +  \mathbfss{V}_{13}^{(0)} \mathbfss{X}_{23}^\dag.
\end{align}
This is \autoref{eq:first-order-visibility}, evaluated explicitly for our 3-antenna model.
Inserting~\autoref{eq:v_1_tot} and~\autoref{eq:v_2_tot} into~\autoref{eq:V_12} directly gives a form of the coupled visibilities that matches \autoref{eq:coupled_terms} and \autoref{eq:coupled_terms_dag}.

\begin{figure*}
    \includegraphics[width=\textwidth]{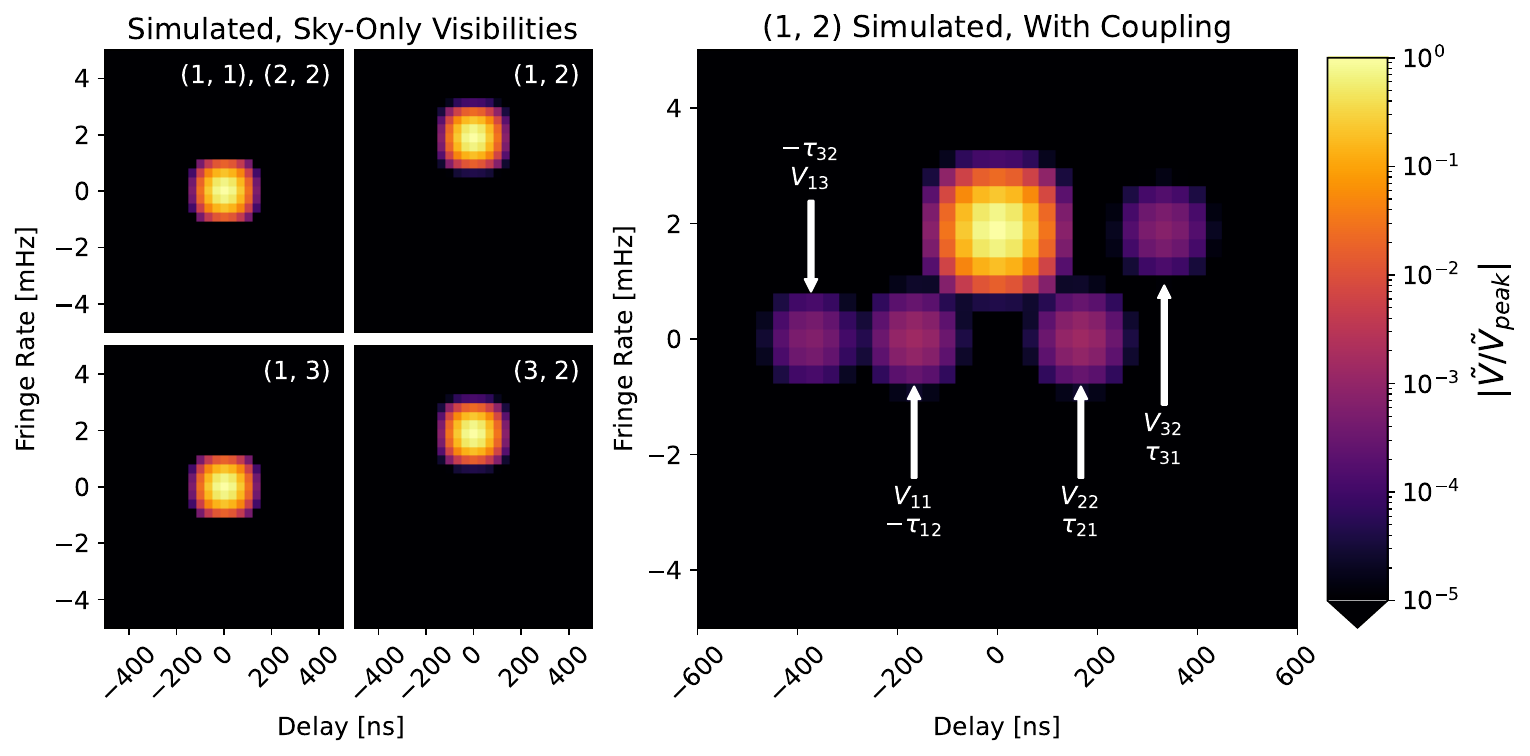}
    \caption{
        \emph{Left}: The sky-only visibilities for our toy visibility model.
        \emph{Right}: A simulation of the first order mutual coupling for baseline (1, 2).
        The coupling pictured here is described by the coupling equation \autoref{eq:3_ant_coupling_eq}.
        The sky signal observed by the baseline resides in the brightest peak centred at zero delay.
        With such a small number of antennas, we can clearly see the power from each coupled visibility residing in separate regions of fringe-rate versus delay space, as predicted by \autoref{eq:tau_f_points}. 
    }
    \label{fig:3_ant_coupled_vis}
\end{figure*}

\subsection{Understanding Delays and Fringe-Rates}
\label{sec:FRATE-DELAY-TRANSFORM}

Now we examine the behavior of mutual coupling in our 3-element array to gain intuition for the model in fringe-rate versus delay space.
Recall that interferometric visibilities are a two-dimensional measurement $V(\nu, t)$.
The delay transform is the Fourier transform along the $\nu$-axis, and the fringe-rate transform is the Fourier transform along the $t$-axis.

Given a frequency spectrum for a single integration, $V(\nu) \big|_t$, we get the delay spectrum when we take the Fourier transform along frequency:
\begin{equation}
    \Tilde{V}(\tau)\Big|_t = \int V(\nu)\Big|_t e^{- 2 \pi i \nu \tau} {\rm d}\nu.
\end{equation}
Although $t$ and $\tau$ both have units of time, they are operating on vastly different timescales and should be thought of independently~\citep{LiuShaw2020}; differences in adjacent times $t$ are of order tens of seconds, while differences in adjacent delays $\tau$ are of order tens of nanoseconds.
Likewise, by taking the Fourier transform of the time stream for a single frequency channel $V(t) \big|_\nu$, we get the fringe-rate transform of the visibility for a single frequency channel:
\begin{equation}
    \Tilde{V}(f)\Big|_\nu = \int V(t)\Big|_\nu e^{- 2 \pi i f t} {\rm d}t.
\end{equation}
Though $\nu$ and $f$ both have units of frequency, it is again important to note that they operate on vastly different scales and should be thought of independently; differences in adjacent frequencies $\nu$ are typically of order hundreds of kHz, while differences in adjacent fringe-rates $f$ are typically of order tens of $\upmu$Hz.
We also point out that~\autoref{eq:3_ant_coupling_eq}, and its generalized counterpart~\autoref{eq:first-order-visibility}, are equations for mutual coupling for visibilities which have not undergone a delay transform.
Since the coefficients are frequency-dependent, but not time-dependent, taking the delay transform of \autoref{eq:first-order-visibility} yields a convolution in delay space.

We evaluate~\autoref{eq:3_ant_coupling_eq} for the array shown in~\autoref{fig:3_ant_array_layout}.
The zeroth-order visibilities are produced using a toy visibility model which is designed to approximately mimic the behavior of foregrounds in fringe-rate versus delay space via
\begin{equation}
    V_{ij} = e^{- \Delta \tau (\nu - \nu_0)^2} e^{i2 \pi \frac{\nu}{c}|\mathbfit{b}_{ij} \times \boldsymbol{\omega}_\oplus|}.
\end{equation}
We use $\Delta \tau$ = 200 ns, which is approximately consistent with observations.
We again use the frequency range 145--165 MHz, and take $\nu_0$ = 155 MHz.
The coupling constants are calculated with a uniform beam.

\autoref{fig:3_ant_coupled_vis} shows the both the contributing zeroth-order visibilities and $V_{12}^{(1)}$.
We see that the zeroth-order visibilities are centred at zero delay and at fringe-rates predicted by \autoref{eq:fringe_rate}.
In the plot of $V_{12}^{(1)}$, the coupled visbilities appear at their original fringe-rate, but are shifted in delay space by the time-of-flight of the coupled radiation.
For this reason, the coupling from the autocorrelations, $V_{11}^{(0)}$ and $V_{22}^{(0)}$, appear symmetrically about zero delay.
These are the points predicted by~\autoref{eq:tau_f_points}.
As we add more antennas, the power from the coupled antennas start to overlap and produce the continuous structures that we see in the HERA data in Section~\ref{sec:EVIDENCE}. 

\bsp	
\label{lastpage}
\end{document}